\documentclass[12pt]{elsarticle}                                          
\usepackage{axodraw4j}
\usepackage{hyperref}
\usepackage{graphicx}                                                         
\usepackage{a41}                                                                
\usepackage{xcolor}                     
\usepackage[rflt]{floatflt}
\usepackage{float}
\usepackage{lscape}
\usepackage{hyperref}
\hypersetup{colorlinks=true, 
linkcolor=blue, filecolor=blue, urlcolor=mygreen}
\usepackage{breakurl} 
\usepackage{times}     %
\usepackage{array}
\usepackage[english]{babel}
\usepackage[latin1]{inputenc}
\usepackage[T1]{fontenc}
\usepackage{ae}
\usepackage{url}
\usepackage{amsmath, amsthm, amssymb}
\usepackage{slashed}

\usepackage{rotating}
\usepackage{graphicx}
\usepackage{comment}
\newcounter{mmacnt}
\def\restartmma{\setcounter{mmacnt}{0}}
\restartmma \catcode`|=\active
\def|#1|{\mathrm{#1}}
\catcode`|=12
\newenvironment{mma}{
\par\smallskip
\catcode`|=\active
\parskip=0pt\parindent=0pt 
\small
\def\In##1\\{%
\def\linebreak{\hfill\break\null\qquad}%
\refstepcounter{mmacnt}
\hangindent=2.5em\hangafter=0
\leavevmode
\llap{\tiny\sffamily In[\arabic{mmacnt}]:=\kern.5em}%
\mathversion{bold}\footnotesize$
\displaystyle##1$\normalsize
\mathversion{normal}\par
 }%
\def\Print##1\\{%
\def\linebreak{\hfill\break}%
\hangindent=2.5em\hangafter=0
\leavevmode ##1\par}%
\def\Out##1\\{%
\def\linebreak{$\hfill\break\null\hfill$}%
\kern\abovedisplayskip\par
\hangindent=2.5em\hangafter=0
\leavevmode
\llap{\tiny\sffamily Out[\arabic{mmacnt}]=\kern.5em}
\footnotesize$\displaystyle##1$
\normalsize\hfill\null\par
\kern\belowdisplayskip
}%
\def\Warning##1##2\\{%
\def\linebreak{\hfill\break}%
\hangindent=2.5em\hangafter=0
\leavevmode
{\scriptsize##1 : ##2}\par}%
}{%
\par\smallskip
}

\usepackage{color}
\newenvironment{fshaded}{%
\MakeFramed {\FrameRestore}
}%
{\endMakeFramed}

\makeatletter
\def\ps@pprintTitle{%
\let\@oddhead\@empty
\let\@evenhead\@empty
\def\@oddfoot{\reset@font\hfil\thepage\hfil}
\let\@evenfoot\@oddfoot
}
\makeatother
\usepackage{tikz}
\usetikzlibrary{matrix}
\allowdisplaybreaks[4]
\newcommand{\n}{\nonumber}
\begin{document}
\begin{frontmatter}
\title{\LARGE
\textbf{One-loop contributions to the decay 
$H\rightarrow \nu_l\bar{\nu}_l\gamma$ in standard model revisited}}
\author[1]{Dzung Tri Tran}
\address[1]{\it University of Science Ho Chi Minh City, $227$ 
Nguyen Van Cu, District $5$, Ho Chi Minh City, Vietnam}
\author[2,3]{Khiem Hong Phan}
\ead{phanhongkhiem@duytan.edu.vn}
\address[2]{\it Institute of Fundamental and Applied Sciences, 
Duy Tan University, Ho Chi Minh City $700000$, Vietnam}
\address[3]{Faculty of Natural Sciences, Duy Tan University, 
Da Nang City $550000$, Vietnam}
\pagestyle{myheadings}
\markright{}
\begin{abstract} 
One-loop contributions to the decay 
$H\rightarrow \nu_l\bar{\nu}_l\gamma$ with $l=e, \mu, \tau$
within standard model framework are revisited in this paper. 
We derive two representations for the form factors in 
this calculation. As a result, the computations are not 
only checked numerically by verifying the ultraviolet 
finiteness of the results but also confirming the ward 
identity of the amplitude. We find that the results are 
good stability with varying ultraviolet cutoff parameters
as well as satisfy the ward identity. In phenomenological 
results, all the 
physical results are examined with the present input 
parameters. Especially, we study the partial decay 
widths for the decay channels in both cases of the detected 
photon and invisible photon. Differential decay widths are 
also generated as a function of energy of final photon. 
\end{abstract}
\end{frontmatter}
\section{Introduction}
\noindent
Today one of the main purposes of experimental program 
at the High Luminosity Large Hadron Collider (HL-LHC) 
\cite{ATLAS:2013hta,CMS:2013xfa} and Future 
Lepton Colliders~\cite{Baer:2013cma} is to measure 
precisely the properties of standard model-like  
(SM-like) Higgs boson. In this program, searching 
for all decay modes of Higgs boson are priority tasks
and have to be probed as precisely as possible.
Because the partial decay widths of Higgs boson provide 
an important information for answering the 
nature of Higgs sector (for our deeper understanding of the 
dynamics of electroweak symmetry breaking). 
In all decay channels of Higgs boson, $H\rightarrow$ 
invisible particles and $H\rightarrow \gamma$ plus 
invisible particles
\cite{Sirunyan:2018owy,Aaboud:2019rtt,Aaboud:2018sfi,
Ngairangbam:2020ksz,Aad:2012re,
Belanger:2013kya,Heikinheimo:2012yd}
are also great of interest by following aspects: 
(i) these processes can be probed at future colliders
\cite{Sirunyan:2018owy,Aaboud:2019rtt,Aaboud:2018sfi,Aad:2012re,
Heikinheimo:2012yd} for testing the standard model (SM) at 
modern energy regions; (ii) many new heavy particles
such as new extra gauge bosons, charged (and neutral) 
scalar particles may exchange in the loop 
diagrams of the aforementioned decay channels. 
Thus, the decay processes could provide a useful 
tool for testing the SM and investigating possible 
new physics.

We know that the precise evaluations 
for the standard model background play a crucial role 
in searching for new physics. It means that the 
theoretical calculations for one-loop contributing 
to $H\rightarrow \nu_l\bar{\nu}_l \gamma$
are necessary. In Ref.~\cite{Sun:2013cba}, one-loop formulas 
for $H\rightarrow \nu_l\bar{\nu}_l \gamma$ 
within SM framework have provided. In view of the 
importance of the decay channels we revisit the 
calculation for the decay processes with extending 
the previous computation and updating numerical predictions 
using the present values of input parameters.
The analytic results for one-loop form factors
are expressed in terms of Passarino-Veltman functions  
(called as PV-functions) which they can be evaluated 
numerically by 
using package {\tt LoopTools}~\cite{Hahn:1998yk}. 
In comparison with the previous work, the 
evaluations in the present paper
are extended by following points. Firstly,
two representations for the form factors are derived 
in this calculation.
As a result, the computations are not 
only checked numerically by verifying the ultraviolet 
finiteness of the results but also confirming the ward 
identity of the amplitude. We find that the results 
are good stability with varying ultraviolet 
cutoff parameters as well as follow the ward identity
of the amplitude. Secondly, we study the partial decay widths 
for the decay channels in both cases of the detected photon 
and invisible photon using the present values of 
input parameters. Lastly, all the PV-functions
are also reduced to scalar one-, two-, three- and four-point 
functions in this paper. In further, we point out that 
one uses the generalised hypergeometric functions 
in \cite{Phan:2018cnz,Phan:2019qee,Phan:2020bde} at general 
space-time dimensions $d$ for scalar one-loop integrals, 
the form factors will be valid in general $d$. 
Therefore, the form factors can be performed  
at higher-power $\varepsilon$-expansion which 
may be taken into account in two-loop and higher-loop 
contributing to these channels. 

The layout of the paper is as follows: In section 2,
detailed calculations for one-loop contributions to 
$H\rightarrow \nu_l\bar{\nu}_l \gamma$ are presented. 
Numerical test for the computations and physical results
for the decay processes are also shown in this sections.
Conclusions and outlook are devoted
in section $3$. In appendices, reduction formulas 
for scalar PV-functions are presented. 
\section{Calculation}    
Detailed calculations for one-loop 
contributions to $H\rightarrow 
\nu_l\bar{\nu}_l \gamma$ 
are presented in this section. 
Two different representations for 
one-loop form factors are derived. 
Numerical tests for the computation and 
physical results for the decay processes are
then shown in the next subsections. 
\subsection{Analytic results}%
\begin{center}
\begin{figure}[hp]
\hspace{1.5cm}
\includegraphics[width=14cm, height=5cm]
{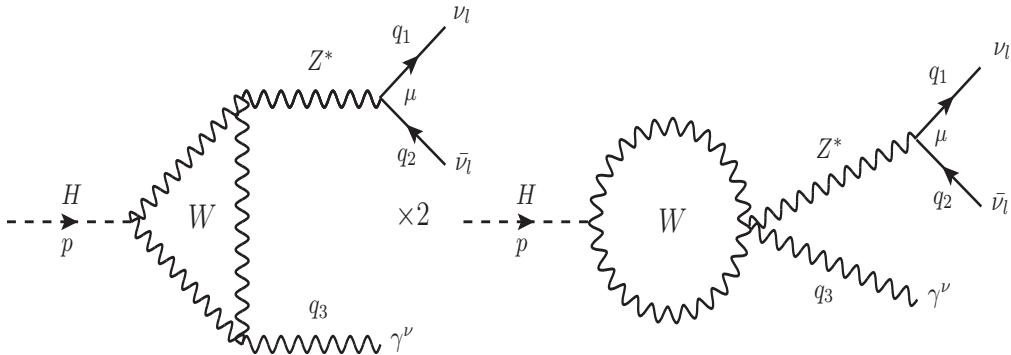}
\caption{\label{trigW} $W$-bosons 
exchanging in the one-loop triangle diagrams  
of $H \rightarrow \nu_l \bar{\nu_l} \gamma $ in 
unitary gauge.}
\end{figure}
\end{center}

\begin{center}
\begin{figure}[ht]
\hspace{1.5cm}
\includegraphics[width=14cm, height=5cm]
{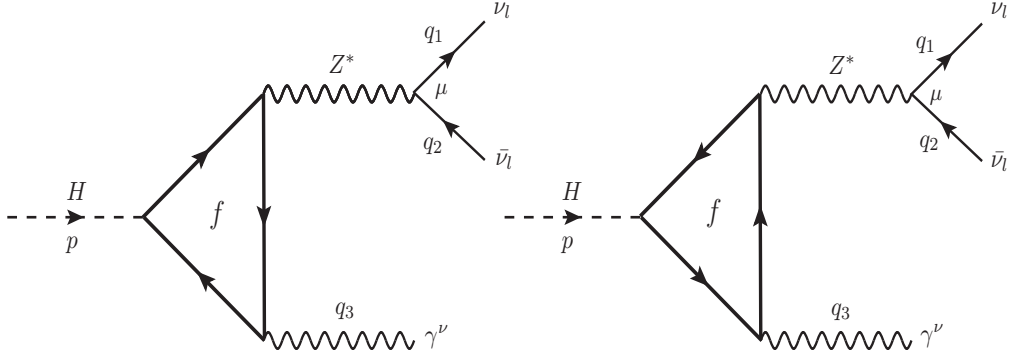}
\caption{\label{trigF} Fermion $f$ exchanging
in the one-loop triangle diagrams  
of $H \rightarrow \nu_l \bar{\nu_l} \gamma $ in 
unitary gauge.}
\end{figure}
\end{center}

\begin{center}
\begin{figure}[ht]
\hspace{0.5cm}
\includegraphics[width=16cm, height=8.5cm]
{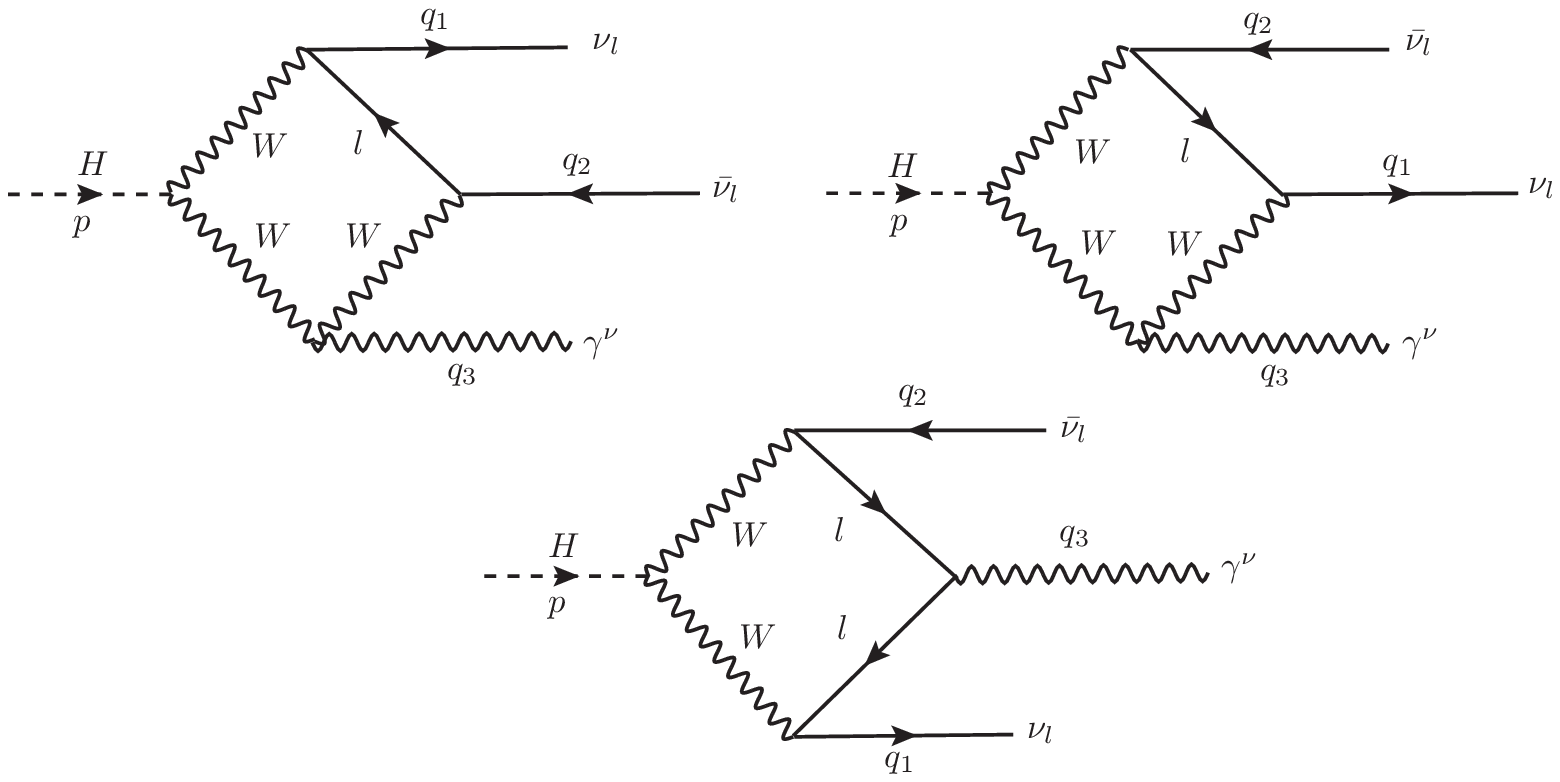}
\caption{\label{boxW} $W$-boson and lepton $l$ 
particles exchanging in the one-loop box diagrams  
of $H \rightarrow \nu_l \bar{\nu_l} \gamma $ in 
unitary gauge.}
\end{figure}
\end{center}

We devote detailed the calculation for the decay channels
$H(p) \rightarrow \nu_{l}(q_1)\bar{\nu}_{l}(q_2) \gamma(q_3)$
in this subsection. Within the standard model framework 
the decay processes consist of $W$ bosons and 
fermions exchanging in one-loop 
triangle diagrams (seen Fig.~\ref{trigW} and Fig.~\ref{trigF} 
respectively) as well as $6$ one-loop box diagrams 
(shown in  Fig.~\ref{boxW}) in unitary gauge. 
\begin{center}
\begin{figure}[ht]
\hspace{0.5cm}
\includegraphics[width=16cm, height=8.5cm]
{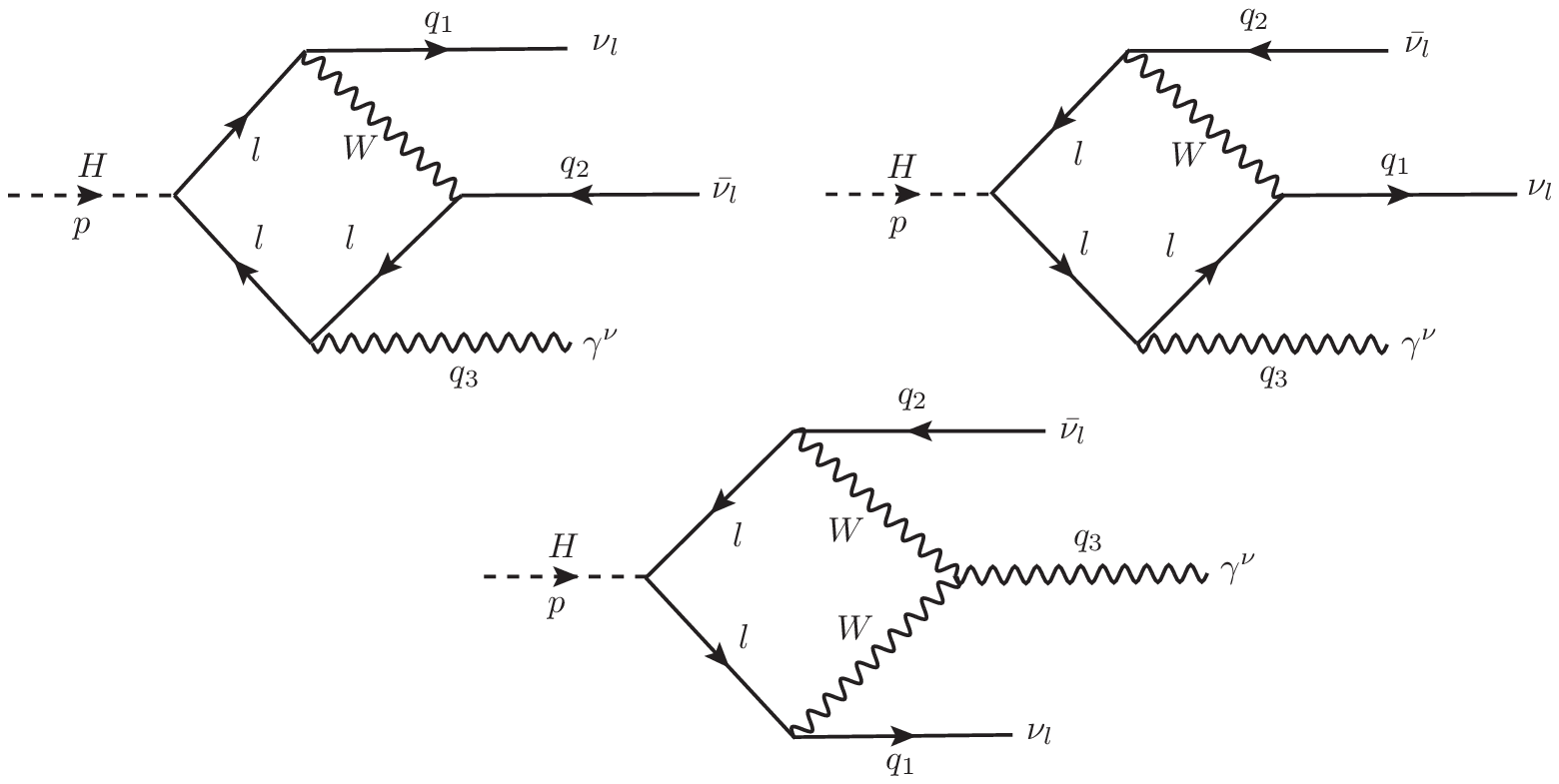}
\caption{\label{boxF} Lepton $l$ and $W$-boson
particles exchanging in the one-loop box diagrams  
of $H \rightarrow \nu_l \bar{\nu_l} \gamma $ in 
unitary gauge.}
\end{figure}
\end{center}
In general, one-loop amplitude of the decay channels 
can be divided as follows:
\begin{eqnarray}
 \mathcal{A} = \mathcal{A}^{(W)}_{\text{tri}}
+ \mathcal{A}^{(f)}_{\text{tri}}
 +\mathcal{A}^{(W)}_{\text{box}} 
 + \mathcal{A}^{(l)}_{\text{box}}.
\end{eqnarray}
In this formula, $\mathcal{A}^{(W)}_{\text{tri}}$ 
presents for the contributions of one-loop 
triangle 
Feynman diagrams with $W$ boson internal lines
(shown in Fig.~\ref{trigW}). The term 
$\mathcal{A}^{(f)}_{\text{tri}}$ is for the 
amplitude of one-loop triangle Feynman 
diagrams with exchanging fermions 
in the loop (seen in Fig.~\ref{trigF}). 
In the same notations, 
$\mathcal{A}^{(W)}_{\text{box}}
(\mathcal{A}^{(l)}_{\text{box}})$ 
is corresponding to the amplitude of 
one-loop box diagrams in Fig.~\ref{boxW} 
(and in Fig.~\ref{boxF} respectively).  

For on-shell external photon, the ward identity is
implied. As a result, we have the following relation:
$q_3^{\nu}\epsilon_{\nu}^*=0$ where $q_3^{\nu}$, 
$\epsilon_{\nu}^*$ are corresponding to
momentum and polarization vector of photon. 
Kinematic invariant variables 
involving to this decay process are concerned: 
\begin{eqnarray}
 p^2 &=& M_H^2, \;  q_1^2= q_2^2=q_3^2=0, \; 
 q_{12} =(q_1+q_2)^2=2 q_1\cdot q_2, 
 \;
 q_{13} = 2q_1\cdot q_3,\; 
 q_{23} = 2q_2\cdot q_3.
\end{eqnarray}
With this definition, one confirms that 
$q_{12} + q_{13} +q_{23}=M_H^2$.

We first write down all Feynman amplitudes for the above 
diagrams. With the help of {\tt Package-X}~\cite{Patel:2015tea}, 
all Dirac traces and Lorentz contractions in $d$ dimensions
are handled. Subsequently, the amplitudes are then 
expressed in terms of tensor one-loop integrals. 
By following tensor reduction for 
one-loop integrals in~\cite{Denner:2005nn}, 
all tensor one-loop integrals are written in terms  
of PV-functions (scalar $PV$-coefficient functions are 
labled as $A_{ijk\cdots}, B_{ijk\cdots}, 
C_{ijk\cdots}$ and $D_{ijk\cdots}$ 
for $i,j, k\cdots = 1,2,3$ as in Ref.~\cite{Denner:2005nn}). 
In this paper, we also reduce the PV-functions to 
scalar one-loop integrals 
(labeled as $A_0, B_0, C_0$ and $D_0$~\cite{Denner:2005nn}). 
Latter, these scalar functions $A_0, B_0, C_0$ and $D_0$
can be evaluated numerically by using {\tt LoopTools}. 
Furthermore, one can use hypergeometric representations
in \cite{Phan:2018cnz,Phan:2019qee,Phan:2020bde} 
for these scalar functions in general $d$. 
\subsubsection{First expressions for form factors} 
We first arrive at one-loop three-point Feynman 
diagrams with $W$ boson internal lines 
(as shown in Fig.~\ref{trigW}). 
The corresponding amplitude for the above Feynman diagrams is 
decomposed into Lorentz invariant structure
as follows: 
\begin{eqnarray}
 \mathcal{A}^{(W)}_{\text{tri}} &=& 
\Big[F_{1}^{(W)} q_3^{\mu}q^{\nu} +  
F_{2}^{(W)}g^{\mu\nu} \Big] 
\bar{u}(q_1)\gamma^{\mu}\; 
P_L v(q_2) \epsilon^{*}_{\nu}(q_3). 
\end{eqnarray}
Where we have used $P_L =\dfrac{1-\gamma_5}{2}$. 
Applying the tensor reduction method in \cite{Denner:2005nn}, 
all form factors in this formula are computed.
Analytic results for the form factors 
are expressed in terms of $B_0$ and $C_0$ functions
as follows:
\begin{eqnarray}
F_{1}^{(W)} &=&
\dfrac{\alpha^2}{2 M_W^3 s_W^3 (d-2)
(M_H^2-q_{12})^2} 
\dfrac{1}{q_{12} -M_Z^2 + i \Gamma_Z M_Z}
\times
\n \\
&&\times 
\Bigg\{
\Big[ 
M_H^2 (2 M_W^2-q_{12})+4 M_W^4 (d-1)-2 M_W^2 q_{12}
\Big]
\times
\n \\
&&\times
\Big[
\Big(
 M_H^2 (d-4)
+(2-d) q_{12}
\Big) 
B_0(M_H^2,M_W^2,M_W^2)
+
2 q_{12} 
B_0(q_{12},M_W^2,M_W^2)
\Big]
\\
&&
+ 4 M_W^2 (M_H^2-q_{12}) 
\Big[
2 M_W^2 (M_H^2 - q_{12}) (2 d-5)
-4 M_W^4 (d-1) \n \\
&&\hspace{2.5cm}
+ q_{12}^2 (d-2)
- M_H^2 q_{12} (d-3)
\Big]
C_0(0,q_{12},M_H^2,M_W^2,M_W^2,M_W^2)
\Bigg\}. \n
\end{eqnarray}
Here $s_W = \sin\theta_W$, $\theta_W$ is weak mixing angle and 
$d = 4-2\varepsilon$ is space-time dimension. In this paper, we 
also compute the form factor $F_2^{(W)}$ which its analytical 
result is shown in the next subsection. In general, $F_2^{(W)}$ 
may take different form with $F_1^{(W)}$. Following the ward 
identity, one confirms that
\begin{eqnarray}
\label{relF12}
F_2^{(W)} = \dfrac{q_{12}-M_H^2}{2} F_1^{(W)} 
\end{eqnarray}
It can be observed the relation 
in Eq.~(\ref{relF12}) analytically 
by performing $\varepsilon$-expansion 
for the form factors. Both the expressions for 
$F_1^{(W)}$ and $F_2^{(W)}$ give same result (seen 
appendix $B$). We also note that the relation 
can be also verified numerically in next subsections.

We now consider Feynman amplitude for one-loop
triangle diagrams with fermion internal lines. 
Following the same procedure, the amplitude is 
casted into the form of
\begin{eqnarray}
 \mathcal{A}^{(f)}_{\text{tri}} &=& 
 \Big[ F_{1}^{(f)} q_3^{\mu}q^{\nu} 
 +  F_{2}^{(f)}g^{\mu\nu}
 \Big] 
 \bar{u}(q_1)\gamma^{\mu}P_L v(q_2) \epsilon^{*}_{\nu}(q_3). 
\end{eqnarray}
All form factors in this equation are evaluated by
using the tensor reduction method in 
\cite{Denner:2005nn}. The results read:
\begin{eqnarray}
F_{1}^{(f)}
&=&
\dfrac{\alpha^2
N_C^f Q_f m_f^2
\Big(
T_3^f - 2 Q_f s_W^2
\Big)}{M_W c_W^2 s_W^3
(d-2)
(M_H^2-q_{12})^2}
\dfrac{1}{ q_{12} -M_Z^2 + i \Gamma_Z M_Z}
\times
\n \\
&& \times
\Bigg\{
\Big[
2 M_H^2 (d-4)- 2 q_{12} (d-2)
\Big] B_0(M_H^2,m_f^2,m_f^2)
+4 q_{12} B_0(q_{12},m_f^2,m_f^2)
\\
&&
+
\Big[
(d-2) (M_H^2-q_{12})^2
-8 m_f^2 (M_H^2-q_{12}) 
\Big] C_0(0,q_{12},M_H^2,m_f^2,m_f^2,m_f^2)
\Bigg\}.
\n
\end{eqnarray}
Here $N_C^f\; (Q_f)$ is number of color and charge quantum 
number of fermion $f$. While $T_{3f}$ is hypercharge 
of fermion $f$. The form factor $F_2^{(f)}$ is shown 
in next subsection. It is important to note that all fermions 
with non-zero masses in the loop are considered in present 
paper. 

We turn our attention to one-loop four-point Feynman 
diagrams contributing to these decay processes. We have 
one-loop box diagrams that Higgs couple directly to 
$W$ bosons (as shown in Fig.~\ref{boxW}). The corresponding 
amplitude is expressed in Lorentz invariant structure as follows: 
\begin{eqnarray}
\label{ampbox-W}
\mathcal{A}^{(W)}_{\text{box}} 
&=&  
\bar{u}(q_1)
\Big\{ 
F_{3}^{(W)} q_1^{\nu} \slashed{q_3}
+  F_{4}^{(W)}q_2^{\nu} \slashed{q_3} 
+ F_{5}^{(W)}
\gamma^{\nu}\Big\}
P_L v(q_2) \epsilon^{*}_{\nu}(q_3). 
\end{eqnarray}
Where $ F_{5}^{(W)}= 
-F_{3}^{(W)} (q_1 \cdot q_3) 
- F_{4}^{(W)} (q_2 \cdot q_3)$. 
Analytic results for all form factors are shown 
in terms of PV-functions as follows:
\begin{eqnarray}
\label{F3W}
F_{3}^{(W)}  &=&
\dfrac{\alpha^2}{2 M_W^5 s_W^3}\times
\n\\
&&\times
\Big\{
(M_H^2 M_W^2+2 M_W^4) 
\Big[
(C_{22} +C_{12} +C_2)(0,q_{12},M_H^2,M_W^2,M_W^2,M_W^2)
\n \\
&&\hspace{3.75cm}
+(C_{22} +C_{12} +C_2)(q_{12},0,M_H^2,M_W^2,M_W^2,M_W^2)
\Big]
\n \\
&&\hspace{0.25cm}
- M_W^2 m_l^2 
\Big[
(C_{22} +C_{12} +C_2)(0,0,q_{13},m_l^2,m_l^2,M_W^2)
\n \\
&&\hspace{2cm}
+(C_{22} +C_{12} +C_2)(0,0,q_{13},m_l^2,M_W^2,M_W^2)
\Big]
\n \\
&&\hspace{0.25cm}
+2 M_W^4 m_l^2 
\Big[
D_0(0,0,0,M_H^2;q_{12},q_{13};M_W^2,m_l^2,M_W^2,M_W^2)
\n \\
&&\hspace{2cm}
+D_0(0,0,0,M_H^2;q_{23},q_{12};M_W^2,M_W^2,m_l^2,M_W^2)
\n \\
&&\hspace{2cm}
+D_0(0,0,0,M_H^2;q_{23},q_{13};M_W^2,m_l^2,m_l^2,M_W^2)
\Big]
\n \\
&&\hspace{0.25cm}
+\Big[
M_W^2 m_l^2 (M_H^2 + 2 M_W^2)
+2 (d-2) M_W^6
\Big]
\times
\n   \\
&&\hspace{0.25cm} \times
\Big[
(D_{33} +D_{23})(0,0,0,M_H^2;q_{12},q_{13};M_W^2,m_l^2,M_W^2,M_W^2)
\n \\
&&\hspace{2cm}
+(D_{33} +D_{23})(0,0,0,M_H^2;q_{23},q_{13};M_W^2,m_l^2,m_l^2,M_W^2)
\n \\
&&\hspace{2cm}
+(D_{33} +D_{23} +D_{13})(0,0,0,M_H^2;q_{23},q_{12};M_W^2,M_W^2,m_l^2,M_W^2)
\Big]
\n \\
&&\hspace{0.25cm}
+ \Big[
M_W^2 m_l^2 (M_H^2 + 2 M_W^2)
+2 (d-4) M_W^6
\Big] 
\times
\n \\
&&\hspace{0.25cm} \times
\Big[D_3(0,0,0,M_H^2;q_{12},q_{13};M_W^2,m_l^2,M_W^2,M_W^2)
\n \\
&&\hspace{2cm}
+D_3(0,0,0,M_H^2;q_{23},q_{12};M_W^2,M_W^2,m_l^2,M_W^2)
\n \\
&&\hspace{2cm}
+D_3(0,0,0,M_H^2;q_{23},q_{13};M_W^2,m_l^2,m_l^2,M_W^2)
\Big]
\n \\
&&\hspace{0.25cm}
+ 4 M_W^4 
\Big[
C_0(0,q_{12},M_H^2,M_W^2,M_W^2,M_W^2)
\n \\
&&\hspace{2cm}
- M_W^2 D_2(0,0,0,M_H^2;q_{12},q_{13};M_W^2,m_l^2,M_W^2,M_W^2)
\Big]
\Big\}.
\end{eqnarray}

Analytic formula for another form factor is shown
\begin{eqnarray}
\label{F4W}
F_{4}^{(W)}  &=&
\dfrac{\alpha^2}{2 M_W^5 s_W^3}
\times
\n \\
&&\hspace{0cm} \times
\Big\{
(M_H^2 M_W^2+2 M_W^4) 
\Big[
(C_{22} +C_{12} +C_2)(0,q_{12},M_H^2,M_W^2,M_W^2,M_W^2)
\n \\
&&\hspace{4.0cm}
+(C_{22} +C_{12} +C_2)(q_{12},0,M_H^2,M_W^2,M_W^2,M_W^2)
\Big]
\n \\
&&\hspace{0.25cm}
- M_W^2 m_l^2 
\Big[
(C_{22} +C_{12} +C_2)(0,0,q_{23},M_W^2,m_l^2,m_l^2)
\n \\
&&\hspace{2.0cm}
+(C_{22} +C_{12} +C_2)(0,0,q_{23},M_W^2,M_W^2,m_l^2)
\Big]
\n \\
&&\hspace{0.25cm}
+2 M_W^4 m_l^2 
\Big[D_0(0,0,0,M_H^2;q_{12},q_{13};M_W^2,m_l^2,M_W^2,M_W^2)
\n \\
&&\hspace{2.0cm}
+D_0(0,0,0,M_H^2;q_{23},q_{12};M_W^2,M_W^2,m_l^2,M_W^2)
\n \\
&&\hspace{2.0cm}
+D_0(0,0,0,M_H^2;q_{23},q_{13};M_W^2,m_l^2,m_l^2,M_W^2)
\Big]
\n \\
&&\hspace{0.25cm}
+\Big[
2 M_W^2 m_l^2 (M_H^2 + 2 M_W^2)
+4 (d-2) M_W^6
\Big] 
\times
\n \\
&&\hspace{0.25cm} \times
\Big[
D_{23}(0,0,0,M_H^2;q_{23},q_{12};M_W^2,M_W^2,m_l^2,M_W^2)
\n \\
&&\hspace{2.cm}
+D_{23}(0,0,0,M_H^2;q_{23},q_{13};M_W^2,m_l^2,m_l^2,M_W^2)
\Big]
\n \\
&&\hspace{0.25cm}
+\Big[
M_W^2 m_l^2 (M_H^2 + 2 M_W^2)
+2 (d-2) M_W^6
\Big]
\times
\\
&&\hspace{0.25cm} \times
\Big[
(D_{33} +D_{23}  +D_{13})(0,0,0,M_H^2;q_{12},q_{13};M_W^2,m_l^2,M_W^2,M_W^2)
\n \\
&&\hspace{1.0cm}
+(D_{33} +D_{22} +D_{13} +D_{12})(0,0,0,M_H^2;q_{23},q_{12};M_W^2,M_W^2,m_l^2,M_W^2)
\n \\
&&\hspace{1.0cm}
+(D_{33} +D_{22} +D_{13} +D_{12})(0,0,0,M_H^2;q_{23},q_{13};M_W^2,m_l^2,m_l^2,M_W^2)
\Big]
\n \\
&&\hspace{0.25cm}
+\Big[
M_W^2 m_l^2 (M_H^2 + 2 M_W^2)
+2 d M_W^6
\Big]
\times
\n \\
&&\hspace{0.25cm} \times
\Big[
D_3(0,0,0,M_H^2;q_{12},q_{13};M_W^2,m_l^2,M_W^2,M_W^2)
\n \\
&&\hspace{2cm}
+(D_3 +D_2)(0,0,0,M_H^2;q_{23},q_{12};M_W^2,M_W^2,m_l^2,M_W^2)
\n \\
&&\hspace{2cm}
+(D_3 +D_2)(0,0,0,M_H^2;q_{23},q_{13};M_W^2,m_l^2,m_l^2,M_W^2)
\Big]
\n \\
&&\hspace{0.25cm}
+4 M_W^6 
\Big[
D_0(0,0,0,M_H^2;q_{23},q_{12};M_W^2,M_W^2,m_l^2,M_W^2)
\n \\
&&\hspace{2cm}
+(D_0 + D_1)(0,0,0,M_H^2;q_{23},q_{13};M_W^2,m_l^2,m_l^2,M_W^2)
\n \\
&&\hspace{2cm}
+(D_2 +D_1 +D_0)(0,0,0,M_H^2;q_{12},q_{13};M_W^2,m_l^2,M_W^2,M_W^2)
\Big]
\n \\
&&\hspace{0.25cm}
+4 M_W^4 C_0(0,q_{12},M_H^2;M_W^2,M_W^2,M_W^2)
\Big\}.
\n
\end{eqnarray}
We have different kinds of $D$-coefficient 
functions relating to tensor reduction for box diagrams appear
in all the above form factors. 
In principle, tensor one-loop box integrals with rank $P=4$
appear in the amplitude of each diagram. However, these 
tensor integrals are cancelled out. As a results, we have up to 
$D_{ij}$ (for $i,j=1,\cdots 3$) functions contributing to 
the aforementioned form factors. 

We finally consider all one-loop box diagrams in 
Fig.~\ref{boxF} in which Higgs couples directly to 
leptons. Amplitude for the diagrams are given as the same form
in Eq.~(\ref{ampbox-W}): 
\begin{eqnarray}
\mathcal{A}^{(l)}_{\text{box}} &=&  
\bar{u}(q_1)
\Big\{F_{3}^{(l)} q_1^{\nu} \slashed{q_3}
+  F_{4}^{(l)}q_2^{\nu} \slashed{q_3} 
+ F_{5}^{(l)} \gamma^{\nu}\Big\}
P_L v(q_2) \epsilon^{*}_{\nu}(q_3)
\end{eqnarray}
with $ F_{5}^{(l)}= 
 -F_{3}^{(l)} (q_1 \cdot q_3) 
- F_{4}^{(l)} (q_2 \cdot q_3)$. 
All related form factors read as follows:
\begin{eqnarray}
\label{F3l}
F_{3}^{(l)}  &=&
\dfrac{\alpha^2}{2 M_W^3 s_W^3} m_l^2
\times
\n \\
&&\times
\Big\{
(C_{22} +C_{12} +C_2)(0,0,q_{13},M_W^2,m_l^2,m_l^2)
+(C_{22} +C_{12} +C_2)(0,0,q_{13},M_W^2,M_W^2,m_l^2)
\n \\
&&\hspace{0.25cm}
+\Big[
(6-d) M_W^2
-m_l^2
\Big]
D_2(0,0,0,M_H^2;q_{12},q_{13};m_l^2,M_W^2,m_l^2,m_l^2)
\n \\
&&\hspace{0.25cm}
+
\Big[ 2 m_l^2 +2 (d-2) M_W^2 \Big]
\Big[
(D_{33}+D_{23})(0,0,0,M_H^2;q_{12},q_{13};m_l^2,M_W^2,m_l^2,m_l^2)
\n \\
&&\hspace{4.0cm}
+(D_{33}+D_{23})(0,0,0,M_H^2;q_{23},q_{13};m_l^2,M_W^2,M_W^2,m_l^2)
\n \\
&&\hspace{4.0cm}
+(D_{33}+D_{23}+D_{13})(0,0,0,M_H^2;q_{23},q_{12};m_l^2,m_l^2,M_W^2,m_l^2)
\Big]
\n \\
&&\hspace{0.25cm}
+
\Big[ (d+2) M_W^2 + m_l^2\Big] 
\Big[ D_3(0,0,0,M_H^2;q_{12},q_{13};m_l^2,M_W^2,m_l^2,m_l^2)
\n \\
&&\hspace{4cm}
+D_3(0,0,0,M_H^2;q_{23},q_{12};m_l^2,m_l^2,M_W^2,m_l^2)
\n \\
&&\hspace{4.cm}
+D_3(0,0,0,M_H^2;q_{23},q_{13};m_l^2,M_W^2,M_W^2,m_l^2)
\Big]
\n \\
&&\hspace{0.25cm}
+2 M_W^2 
\Big[
D_0(0,0,0,M_H^2;q_{12},q_{13};m_l^2,M_W^2,m_l^2,m_l^2)
\n \\
&&\hspace{4cm}
+D_0(0,0,0,M_H^2;q_{23},q_{12};m_l^2,m_l^2,M_W^2,m_l^2)
\n \\
&&\hspace{4cm}+D_0(0,0,0,M_H^2;q_{23},q_{13};m_l^2,M_W^2,M_W^2,m_l^2)
\Big]
\Big\}
\end{eqnarray}
and 
\begin{eqnarray}
\label{F4l}
F_{4}^{(l)}  &=&
\dfrac{\alpha^2}{2 M_W^3 s_W^3} m_l^2
\times
\n \\
&&\times
\Big\{
(C_{22} +C_{12} +C_2)(0,0,q_{23};m_l^2,m_l^2,M_W^2)
+(C_{22} +C_{12} +C_2)(0,0,q_{23};m_l^2,M_W^2,M_W^2)
\n \\
&&\hspace{0.25cm}
+
\Big[
4 m_l^2
+4 (d-2) M_W^2
\Big]
\Big[
D_{23}(0,0,0,M_H^2;q_{23},q_{12};m_l^2,m_l^2,M_W^2,m_l^2)
\n \\
&&\hspace{4.5cm}
+D_{23}(0,0,0,M_H^2;q_{23},q_{13};m_l^2,M_W^2,M_W^2,m_l^2)
\Big]
\n \\
&&\hspace{0.25cm}
+\Big[
2 m_l^2
+2 (d-2) M_W^2
\Big] 
\Big[
(D_{33}+D_{23}+D_{13})(0,0,0,M_H^2;q_{12},q_{13};m_l^2,M_W^2,m_l^2,m_l^2)
\n \\
&&\hspace{3.cm}
+(D_{33}+D_{22}+D_{13}+D_{12})(0,0,0,M_H^2;q_{23},q_{12};m_l^2,m_l^2,M_W^2,m_l^2)
\n \\
&&\hspace{3.cm}
+(D_{33}+D_{22}+D_{13}+D_{12})(0,0,0,M_H^2;q_{23},q_{13};m_l^2,M_W^2,M_W^2,m_l^2)
\Big]
\n \\
&&\hspace{0.25cm}
+
\Big[ m_l^2+(d-4) M_W^2 \Big] 
\Big[
D_0(0,0,0,M_H^2;q_{12},q_{13};m_l^2,M_W^2,m_l^2,m_l^2)
\n \\
&&\hspace{4cm}
+D_0(0,0,0,M_H^2;q_{23},q_{12};m_l^2,m_l^2,M_W^2,m_l^2)
\n \\
&&\hspace{4cm}
+D_0(0,0,0,M_H^2;q_{23},q_{13};m_l^2,M_W^2,M_W^2,m_l^2)
\Big]
\n \\
&&\hspace{0.25cm}
+\Big[3 m_l^2 + (3 d - 10) M_W^2 \Big] 
\Big[
D_3(0,0,0,M_H^2;q_{12},q_{13};m_l^2,M_W^2,m_l^2,m_l^2)
\n \\
&&\hspace{4.cm}
+(D_3+D_2)(0,0,0,M_H^2;q_{23},q_{12};m_l^2,m_l^2,M_W^2,m_l^2)
\n \\
&&\hspace{4cm}
+(D_3+D_2)(0,0,0,M_H^2;q_{23},q_{13};m_l^2,M_W^2,M_W^2,m_l^2)
\Big]
\n \\
&&\hspace{0.25cm}
+\Big[ m_l^2 +(d-6) M_W^2 \Big]
\Big[
D_1(0,0,0,M_H^2;q_{23},q_{13};m_l^2,M_W^2,M_W^2,m_l^2)
\n \\
&&\hspace{4cm}
+( D_2 +D_1)
(0,0,0,M_H^2;q_{12},q_{13};m_l^2,M_W^2,m_l^2,m_l^2)
\Big]
\Big\}.
\end{eqnarray}
We find that this contribution is much smaller
than other artributions because it is 
proportional $m_l^2$. It is enough to 
take $\tau$-lepton for this contribution. 
\subsubsection{Second expression for form factors} 
Another expression for all form factors appearing 
in this paper are also presented in this subsection.
\begin{eqnarray}
F_{2}^{(W)}  &=&
\dfrac{\alpha^2}{2 M_W^3 s_W^3
(q_{12} -M_Z^2 + i \Gamma_Z M_Z)}
\Bigg\{
\dfrac{q_{12}}{(d-1)}
\Big[
2 M_W^2 B_0(0,M_W^2,M_W^2)
-
(d-2) A_0(M_W^2)
\Big]
\n \\
&&
+
\dfrac{q_{12} }{(d-2) (M_H^2-q_{12})}
\Big[M_H^2 (q_{12}-2 M_W^2) 
- 4 (d-1) M_W^4+2 M_W^2 q_{12}\Big]
B_0(q_{12},M_W^2,M_W^2)
\n \\
&&
+
\dfrac{1}{2 (2-d) (M_H^2-q_{12})}
\Big\{
(d-4) M_H^4 (2 M_W^2-q_{12})
-2 M_W^2 q_{12} (d-2) \Big[2 M_W^2 (d-1)-q_{12}\Big]
\n \\
&&
+M_H^2 
\Big[4 M_W^4 (d-1)(d-4) - 
4 M_W^2 q_{12} (d-3)+(d-2) q_{12}^2\Big] \Big\}
B_0(M_H^2,M_W^2,M_W^2)
\n \\
&&
+
\dfrac{2 M_W^2 }{(d-2)}
\Big\{
M_H^2 \Big[(10-4 d) M_W^2+(d-3) q_{12}\Big]
\\
&&\hspace{0cm}
+4 M_W^4 (d-1)
+2 M_W^2 q_{12} (2 d-5)-(d-2) q_{12}^2
\Big\}
C_0(0,q_{12},M_H^2,M_W^2,M_W^2,M_W^2)
\Bigg\}. \n
\end{eqnarray}
For fermion particles exchanging in one-loop
triangle Feynman diagrams, we also have the following
form factor:
\begin{eqnarray}
F_{2}^{(f)}
&=&
\dfrac{\alpha^2 \, N_C^f Q_f m_f^2
\Big(
T_3^f - 2 Q_f s_W^2
\Big)}{2 M_W c_W^2 s_W^3 (d-2) (M_H^2-q_{12})
(q_{12} -M_Z^2 + i \Gamma_Z M_Z)}
\times
\\
&&\times 
\Bigg\{
\Big[
4 M_H^2
-2 (d-2) (M_H^2-q_{12}) 
\Big] B_0(M_H^2,m_f^2,m_f^2)
-4 q_{12} B_0(q_{12},m_f^2,m_f^2)
\n \\
&&
\hspace{0.5cm} 
+ (M_H^2-q_{12}) \Big[
8 m_f^2
- (d-2) (M_H^2-q_{12})
\Big]
C_0(0,q_{12},M_H^2,m_f^2,m_f^2,m_f^2)
\Bigg\}.
\n
\end{eqnarray}
Form factor $F_{5}^{(W)}$ of the box diagrams in Fig.~(\ref{boxW})
is expressed: 
\begin{eqnarray}
F_{5}^{(W)} &=&
\dfrac{\alpha^2}{4 M_W^5 s_W^3}
\times
\n\\
&&\times
\Big\{
2 M_W^2 A_0(M_W^2)
-4 M_W^2 B_{00}(0,M_W^2,M_W^2)
-M_H^2 M_W^2 B_0(M_H^2,M_W^2,M_W^2)
\n \\
&&\hspace{0.25cm}
+ M_W^2 (2 M_W^2 + m_l^2) B_0(0,M_W^2,m_l^2)
+ M_W^2 (m_l^2-4 M_W^2) B_0(q_{13},m_l^2,M_W^2)
\n \\
&&\hspace{0.25cm}
- M_W^2 m_l^2 q_{23} 
(C_2+C_1+C_0)
(0,0,q_{23},M_W^2,m_l^2,m_l^2)
\n \\
&&\hspace{0.25cm}
-2 M_W^2 m_l^2 
\Big[
C_{00}(0,0,q_{13},m_l^2,m_l^2,M_W^2)
+C_{00}(0,0,q_{13},m_l^2,M_W^2,M_W^2)
\n \\
&&\hspace{2cm}
+C_{00}(0,0,q_{23},M_W^2,m_l^2,m_l^2)
+C_{00}(0,0,q_{23},M_W^2,M_W^2,m_l^2)
\Big]
\n \\
&&\hspace{0.25cm}
+
\Big[
2 M_W^6 (1-d)
-2 M_H^2 M_W^4
-M_H^2 m_l^2 M_W^2
\Big]
C_0(0,q_{13},M_H^2,M_W^2,m_l^2,M_W^2)
\n \\
&&\hspace{0.25cm}
+2 M_W^4 
\Big\{
\Big[
(q_{13}-3 M_H^2) 
C_2
+(q_{13}-M_H^2) 
C_1
\Big]
(0,q_{13},M_H^2,M_W^2,m_l^2,M_W^2)
\n \\
&&\hspace{2cm}
+2(q_{12} - M_H^2) 
C_0(0,q_{12},M_H^2,M_W^2,M_W^2,M_W^2)
\Big\}
\n \\
&&\hspace{0.25cm}
+2  M_W^2 (2 M_W^2 + M_H^2) 
\Big[
C_{00}(0,q_{12},M_H^2;M_W^2,M_W^2,M_W^2)
\n \\
&&\hspace{5.cm}
+C_{00}(q_{12},0,M_H^2;M_W^2,M_W^2,M_W^2)
\Big]
\n \\
&&\hspace{0.25cm}
-\Big[
2 M_W^4 q_{13} (2 M_W^2 + m_l^2) 
+2  M_W^4 q_{23} ( 4 M_W^2 + m_l^2 )
\Big]
\times
\\
&&\hspace{5.25cm} \times
D_0(0,0,0,M_H^2;q_{12},q_{13};M_W^2,m_l^2,M_W^2,M_W^2)
\n \\
&&\hspace{0.25cm}
-\Big[
2 M_W^4 q_{13} m_l^2
+2 M_W^4 q_{23} (2 M_W^2 + m_l^2)
\Big] 
\times
\n \\
&&\hspace{5.25cm} \times
D_0(0,0,0,M_H^2;q_{23},q_{12};M_W^2,M_W^2,m_l^2,M_W^2)
\n \\
&&\hspace{0.25cm}
+\Big\{
M_W^2 q_{23} 
\Big[
M_H^2 m_l^2 
+ 2 M_W^4 (d-6)
\Big]
-2 M_W^4 q_{13} m_l^2
\Big\}
\times
\n \\
&&\hspace{5.25cm} \times
D_0(0,0,0,M_H^2;q_{23},q_{13};M_W^2,m_l^2,m_l^2,M_W^2)
\n \\
&&\hspace{0.25cm}
+ M_W^2 q_{23} \Big[
2 M_W^4 (d-6)
+ m_l^2 (2 M_W^2+M_H^2)
\Big]
\times
\n \\
&&\hspace{3.cm} \times
(D_3 +D_2+D_1)(0,0,0,M_H^2;q_{23},q_{13};M_W^2,m_l^2,m_l^2,M_W^2)
\n \\
&&\hspace{0.25cm}
+\Big[
4 M_W^6 (d-2)
+2 M_W^2 m_l^2 (M_H^2 + 2 M_W^2)
\Big] 
\times
\n \\
&&\hspace{2.cm} \times
\Big[
D_{00}(0,0,0,M_H^2;q_{12},q_{13};M_W^2,m_l^2,M_W^2,M_W^2)
\n \\
&&\hspace{2cm}
+D_{00}(0,0,0,M_H^2;q_{23},q_{12};M_W^2,M_W^2,m_l^2,M_W^2)
\n \\
&&\hspace{2cm}
+D_{00}(0,0,0,M_H^2;q_{23},q_{13};M_W^2,m_l^2,m_l^2,M_W^2)
\Big]
\n \\
&&\hspace{0.25cm}
-8 M_W^6 q_{23} 
\Big[
(D_3 + D_2 + D_1)
(0,0,0,M_H^2;q_{12},q_{13};M_W^2,m_l^2,M_W^2,M_W^2) 
\n \\
&&\hspace{2cm}
+ (D_3 + D_2)
(0,0,0,M_H^2;q_{23},q_{12};M_W^2,M_W^2,m_l^2,M_W^2) 
\Big]
\Big\}.
 \n
\end{eqnarray}

Similarly, form factor $F_{5}^{(l)}$ of one-loop 
box diagrams in Fig.~\ref{boxF} is as follows:
\begin{eqnarray}
F_{5}^{(l)} &=&
\dfrac{\alpha^2}{4 M_W^3 s_W^3} m_l^2
\times
\n \\
&&\times
\Big\{
B_0(M_H^2,m_l^2,m_l^2)
-3 B_0(q_{13},M_W^2,m_l^2)
-q_{23} C_2(0,0,q_{23},m_l^2,m_l^2,M_W^2)
\n \\
&&\hspace{0.25cm}
+\Big[
(q_{13} - 3 M_H^2) C_2 
+(q_{13} - M_H^2) C_1
\Big]
(0,q_{13},M_H^2,m_l^2,M_W^2,m_l^2)
\n \\
&&\hspace{0.25cm}
+\Big[
(5-2d) M_W^2
-M_H^2
-3 m_l^2
\Big]
C_0(0,q_{13},M_H^2,m_l^2,M_W^2,m_l^2)
\n \\
&&\hspace{0.25cm}
+2 C_{00}(0,0,q_{13},M_W^2,m_l^2,m_l^2)
+2 C_{00}(0,0,q_{13},M_W^2,M_W^2,m_l^2)
\n \\
&&\hspace{0.25cm}
+2 C_{00}(0,0,q_{23},m_l^2,m_l^2,M_W^2)
+2 C_{00}(0,0,q_{23},m_l^2,M_W^2,M_W^2)
\n \\
&&\hspace{0.25cm}
+\Big[
4 M_W^2 (d-2)
+4 m_l^2
\Big] 
\Big[
D_{00}(0,0,0,M_H^2;q_{12},q_{13};m_l^2,M_W^2,m_l^2,m_l^2)
\n \\
&&\hspace{4.25cm}
+D_{00}(0,0,0,M_H^2;q_{23},q_{12};m_l^2,m_l^2,M_W^2,m_l^2)
\n \\
&&\hspace{4.25cm}
+D_{00}(0,0,0,M_H^2;q_{23},q_{13};m_l^2,M_W^2,M_W^2,m_l^2)
\Big]
\\
&&\hspace{0.25cm}
+ q_{13} \Big[
(4-d) M_W^2 - m_l^2 \Big]
D_0(0,0,0,M_H^2;q_{12},q_{13};m_l^2,M_W^2,m_l^2,m_l^2)
\n\\
&&\hspace{0.25cm}
+ q_{23}
\Big[ (10 -2 d) M_W^2 -2 m_l^2 \Big]
D_0(0,0,0,M_H^2;q_{12},q_{13};m_l^2,M_W^2,m_l^2,m_l^2)
\n\\
&&\hspace{0.25cm}
+\Big[
(4-d) M_W^2 q_{23}  
-m_l^2q_{23}
-2 M_W^2 q_{13}
\Big]
D_0(0,0,0,M_H^2;q_{23},q_{12};m_l^2,m_l^2,M_W^2,m_l^2)
\n \\
&&\hspace{0.25cm}
+ 2 q_{23} \Big[
(6-d) M_W^2 
-m_l^2 
\Big]
\times
\n \\
&&\hspace{2.cm} \times 
\Big[
(D_3+D_2+D_1)
(0,0,0,M_H^2;q_{12},q_{13};m_l^2,M_W^2,m_l^2,m_l^2)
\n \\
&&\hspace{2.25cm}
+(D_3+D_2)
(0,0,0,M_H^2;q_{23},q_{12};m_l^2,m_l^2,M_W^2,m_l^2)
\Big]
\n \\
&&\hspace{0.25cm}
+8 M_W^2 q_{23} 
(D_3+D_2+D_1)
(0,0,0,M_H^2;q_{23},q_{13};m_l^2,M_W^2,M_W^2,m_l^2)
\n \\
&&\hspace{0.25cm}
+2 M_W^2 (3 q_{23}- q_{13}) 
D_0(0,0,0,M_H^2;q_{23},q_{13};m_l^2,M_W^2,M_W^2,m_l^2)
\Big\}.
\n
\end{eqnarray}
Having the form factors, we are going to calculate 
the decay widths. Differential decay width is derived 
in detail in Appendix $C$. 
Several decay width formulas are presented in terms of 
the above form factors. In this subsection, we present one of
the differential decay width expression which written in terms of
$F_1$, $F_3$ and $F_4$. The formula is as follows: 
\begin{eqnarray}
\label{decay1}
 \dfrac{d^2\Gamma(H\rightarrow 
\nu_l\bar{\nu}_l  \gamma)}{dq_{12}dq_{13}}
 &=& \dfrac{q_{12}}{512\pi^3 M_H^3}
\Big\{
\big(|F_1|^2 + |F_3|^2 + 2\, \text{Re} [F_1 F_3^*]\big) q_{13}^2 
\\
&&\hspace{3.5cm} + 
\big(|F_1|^2 + |F_4|^2 + 2\, \text{Re} [F_1 F_4^*]\big) q_{23}^2
\Big\}.\n
\end{eqnarray}
Where 
\begin{eqnarray}
 F_{i} &=& \sum\limits_{f}F_{i}^{(f)} + F_{i}^{(W)},\quad 
 F_{j} = \sum\limits_{l=e, \mu,\tau}F_{j}^{(l)} + F_{j}^{(W)}
\end{eqnarray}
for $i=1,2$ and $j=3,4,5$. 
Where the integration region is
\begin{eqnarray}
0 \leq q_{12} \leq M_H^2,\quad 0 \leq q_{13} \leq M_H^2-q_{12}. 
\end{eqnarray}
Other forms for the differential 
decay width which expressed in terms of 
$F_2$, $F_3$ (or $F_4$) and $F_5$  are shown in appendix $C$. 
It is important to
note that all form factors appearing in this paper
are used for computing the decay widths.

In appendix $B$, all reduction formulas for 
PV-functions expressing in terms of scalar one-loop 
integrals $A_0, B_0, C_0$ and $D_0$ are derived in space-time 
dimensions $d$. Analytical results for these scalar integrals
are well-known in $d= 4-2\varepsilon$ at $\varepsilon^0$-expansion
and they can be evaluated numerically by using {\tt LoopTools}
~\cite{Hahn:1998yk}.

We are going to end up the analytical 
calculation with interesting point for future prospect.  
Recently, we have derived hypergeometric representations
for scalar one-loop in general space-time dimensions $d$ in 
Refs.~\cite{Phan:2018cnz,Phan:2019qee,Phan:2020bde}. 
By expressing scalar one-loop 
integrals $A_0, B_0, C_0$ and $D_0$ in terms of hypergeometric
functions in Refs.~\cite{Phan:2018cnz,Phan:2019qee,Phan:2020bde},
we support that all form factors in this paper will 
be valid in general $d$. 
Subsequently, they can be performed at higher-power of 
$\varepsilon$-expansion which these terms may be taken 
into account in general framework for 
two-loop and higher-loop contributing to the aforementioned 
channels.
\subsection{Phenomenological results}         
All the physical results of the decay channels 
are examined with the present input parameters in 
\cite{Zyla:2020zbs}. In detail, we use following input parameters:
$\alpha = 1/137.035 999 084$, $M_Z = 91.1876$ GeV, 
$\Gamma_Z  = 2.4952$ GeV, $M_W = 80.379$ GeV, $M_H =125.1$ GeV,
$m_{\tau} = 1.77686$ GeV, $m_t= 172.76$ GeV, 
$m_b= 4.18$ GeV, $m_s = 0.93$ GeV and  $m_c = 1.27$ GeV.  

Before we are going to discuss phenomenological results
for these processes. First, numerical checks for the computations
are performed. The results must be independent of 
ultraviolet cutoff 
($C_{UV}=1/\varepsilon$) and 
$\mu^2$ parameters ($\mu^2$ plays role of a renormalization scale 
\cite{Hahn:1998yk}). We take the 
form factors $F_1^{(W)}$, $F_3^{(W)} $ and $F_4^{(W)}$ as typical 
examples. Numerical results are shown at arbitrary 
sampling point in physical region (seen Tables \ref{NumF1},
\ref{NumF2}, \ref{NumF3}).  
\begin{table}[h]
\begin{center}
\begin{tabular}{lll}  \hline \\
$\textbf{Diagrams} /(C_{UV}, \mu^2)$
& $(0, 1)$ & $(10^{5}, 10^{7})$   \\ \\ \hline \hline\\
diag.~1& $3.822254271421409 \cdot 10^{-9}$ & $-0.000025848850122985293$ \\ 
& + $1.1328167325068205 \cdot 10^{-10} \, i$ &  $-7.660926734863565 \cdot 10^{-7} \, i$  
  \\ \hline\\
diag.~2 & $-4.284907193808991 \cdot 10^{-9}$ & $0.00005170105984731962$  \\ 
  &  $-1.2699350230775137 \cdot 10^{-10} \, i$ & + $1.5322849168169064 \cdot 10^{-6} \, i$  
  \\ \hline\\
$2\times$ diag.~1 $+$ diag.~2 & $3.3596013490338273 \cdot 10^{-9}$ & $3.3596013490338273 \cdot 10^{-9}$  \\   
& + $9.956984419361279 \cdot 10^{-11} \, i$ & + $9.956984419361279 \cdot 10^{-11} \, i$  
\\ \hline\hline
\end{tabular}
\caption{\label{NumF1} Numerical checks for $F_1^{(W)}$ 
of triangle diagrams with exchanging $W$ bosons}
\end{center}
\end{table}
\begin{table}[h]
\begin{center}
\begin{tabular}{lll}  \hline \\
$\textbf{Diagrams} /(C_{UV}, \mu^2)$
& $(0, 1)$ & $(10^5, 10^7)$   \\ \\ \hline \hline\\
diag.~1  & $-1.510623096811763 \cdot 10^{-10}$ & 
$-1.510623096811763 \cdot 10^{-10}$ 
\\ 
&  $-1.703281388817785 \cdot 10^{-24} \, i$ &  $-1.703281388817785 \cdot 10^{-24} \, i$
  \\ \hline\\
diag.~2  & $6.682548276674295 \cdot 10^{-10}$ 
& $6.682548276674295 \cdot 10^{-10}$ 
\\ 
&  + $6.582748337547062 \cdot 10^{-10} \, i$ & + $6.582748337547062 \cdot 10^{-10} \, i$
  \\ \hline\\
diag.~3  & $-1.5802922737417574 \cdot 10^{-10}$ 
& $-1.5802922737417574 \cdot 10^{-10}$
\\ 
&  + $1.680875383453354 \cdot 10^{-10} \, i$ & + $1.680875383453354 \cdot 10^{-10} \, i$
  \\ \hline\\
Sum & $3.591632904379816 \cdot 10^{-10}$ 
& $3.591632904379816 \cdot 10^{-10}$  \\   
  & + $8.263623719173174 \cdot 10^{-10} \, i$ & + $8.263623719173174 \cdot 10^{-10} \, i$  \\
   \hline\hline
\end{tabular}
\caption{\label{NumF2} Numerical checks for 
$F_3^{(W)}$ box diagrams with exchanging $W$ bosons}
\end{center}
\end{table}
\begin{table}[ht]
\begin{center}
\begin{tabular}{lll}  \hline \\
$\textbf{Diagrams} /(C_{UV}, \mu^2)$
& $(0, 1)$ & $(10^5, 10^7)$   \\ \\ \hline \hline\\
diag.~1  & $-2.8890520180530153 \cdot 10^{-10}$ & $-2.8890520180530153 \cdot 10^{-10}$
\\ 
& + $3.09465553621708 \cdot 10^{-25} \, i$ & + $3.09465553621708 \cdot 10^{-25} \, i$
  \\ \hline\\
diag.~2  & $5.932051711362282 \cdot 10^{-10}$ & $5.932051711362282 \cdot 10^{-10}$
\\ 
& + $9.380790178780355 \cdot 10^{-10} \, i$ & + $9.380790178780355 \cdot 10^{-10} \, i$
  \\ \hline\\
diag.~3  & $7.503951368583933 \cdot 10^{-11}$ & $7.503951368583933 \cdot 10^{-11}$
\\ 
& + $3.476774032077318 \cdot 10^{-10} \, i$ & + $3.476774032077318 \cdot 10^{-10} \, i$
  \\ \hline\\
Sum & $3.793394830167515 \cdot 10^{-10}$ & $3.793394830167515 \cdot 10^{-10}$
\\ 
& + $1.2857564210858306 \cdot 10^{-9} \, i$ & + $1.2857564210858306 \cdot 10^{-9} \, i$
  \\ 
   \hline\hline
\end{tabular}
\caption{\label{NumF3} Numerical checks for $F_4^{(W)}$ box diagrams 
with exchanging $W$ bosons}
\end{center}
\end{table}

After verifying the ultraviolet finiteness  and $\mu^2$ independent 
of the form factors, we perform a further test for the results 
which is to the ward identity check for the amplitudes. As we 
mentioned in previous sections, two representations
for form factors are derived in this paper. Their relations are tested numerically. 
Numerical results are shown at arbitrary sampling point in physical region: 
\begin{eqnarray}
 F_{1}^{(W)} 
 &=& \left(
 \dfrac{2}{q_{12} -M_H^2} \right) F_{2}^{(W)} =
  4.417241820666953 \cdot 10^{-8} + 5.8717660992932236\cdot 10^{-8} \, i,
 \n 
 \\
 &&\\
 &&\n\\
 F_{1}^{(\tau)} 
 &=& \left(
 \dfrac{2}{q_{12} -M_H^2} \right) F_{2}^{(\tau)} =
 -2.906788516027605 \cdot 10^{-12} - 1.660128209352433 \cdot 10^{-12} \, i,
 \n \\
&&\\
\label{F5W}
F_{5}^{(W)}&=& 
 -F_{3}^{(W)} (q_1 \cdot q_3) 
- F_{4}^{(W)} (q_2 \cdot q_3)=
\n\\
&&\hspace{3.3cm}=
-8.31386693767833 \cdot 10^{-7} + 2.076359788449464 \cdot 10^{-20} \, i,
\\
\label{F5l}
&&\n\\
F_{5}^{(\tau)}&=& 
-F_{3}^{(\tau)} (q_1 \cdot q_3) 
- F_{4}^{(\tau)} (q_2 \cdot q_3) =
\n\\
&&\hspace{3.3cm} =
9.43776046706387 \cdot 10^{-9} - 3.509321200912556 \cdot 10^{-9} \, i.
\end{eqnarray}
We find that the results 
are good stability when varying $C_{UV}, \mu^2$ parameters
as well as following the ward identity checks. 

The computations are confirmed successfully by numerical checks. 
The phenomenological results for the decay channels are studied. 
In present paper both event of  Higgs decay to invisible particles
and a photon plus invisible particles are computed.
We first arrive at the case of photon that may be tested or not. 
In this case, we do not apply any energy cut for final photon. 
The partial decay width for $H\rightarrow \nu_l\bar{\nu}_l\gamma$
in which $l$ can be one of  $e, \mu$ and $\tau$ is presented: 

\begin{eqnarray}
\Gamma_{H\rightarrow \gamma \nu_l \bar{\nu}_l}^{\text{tot}} 
&=& 0.471109(5) \, \text{KeV},
\n \\
\n \\
\Gamma_{H\rightarrow \gamma \nu_l \bar{\nu}_l}^{\text{Trig}} 
&=& 0.464686(1) \, \text{KeV},
\n\\
\n \\
\Gamma_{H\rightarrow \gamma \nu_l \bar{\nu}_l}^{\text{Trig}\times
\text{W-box} } &=& 0.006064(4)\, \text{KeV},
\n\\
\n \\
\Gamma_{H\rightarrow \gamma \nu_l \bar{\nu}_l}^{\text{Trig}\times
\text{lepton-box} } &=& 0.000042(5) \, \text{KeV}.
\n
\end{eqnarray}
$\Gamma_{H\rightarrow \gamma \nu_l \bar{\nu}_l}^{\text{tot}}$ is 
for the case of taking all Feynman diagrams. It gives a perfect
agreement with the result in~\cite{Sun:2013cba}. 
$\Gamma_{H\rightarrow \gamma \nu_l \bar{\nu}_l}^{\text{Trig}}$ 
is noted for the contributions of all triangle diagrams. 
This gives dominant contributions in comparison with other
parts. The contributions of the interference 
between three-point diagrams and box diagrams 
$\Gamma_{H\rightarrow \gamma \nu_l \bar{\nu}_l}^{\text{Trig}\times \text{W-box} }$ 
(box diagrams in Fig.~\ref{boxW}), 
$\Gamma_{H\rightarrow \gamma \nu_l \bar{\nu}_l}^{\text{Trig}
\times\text{lepton-box}}$ (for box diagrams in Fig.~\ref{boxF})
are shown in these equations.  These contributions 
are much smaller than the results of triangle 
diagrams.

We also concern the case of photon that can be tested. 
In this case, we apply energy cuts for final photon.
The results are shown in Table \ref{testPHOTON}. 
\begin{table}[h]
\centering
\begin{tabular}{lccc}  \hline \\
$\Gamma$ [KeV]/$E_{\gamma}^{\text{cut}}$ [GeV]
& $5$    & $30$      & $50$ \\ \hline\hline \\
$\Gamma_{H\rightarrow \gamma \nu_l \bar{\nu}_l}^{\text{Trig}} $
&$0.464684(7)$ & $0.171761(3)$ & $0.004632(7)$\\ \hline \\
$\Gamma_{H\rightarrow \gamma \nu_l \bar{\nu}_l}^{\text{Trig}\times
\text{W-box} }$
&$0.006064(2)$ & $0.005928(6)$ & $0.001541(7)$\\ \hline \\
$\Gamma_{H\rightarrow \gamma \nu_l \bar{\nu}_l}^{\text{Trig}\times 
\text{lepton-box} }$
&$0.000042(4)$ & $0.000006(2)$ & $0.000000(1)$ \\ \hline \\
$\Gamma_{H\rightarrow \gamma \nu_l \bar{\nu}_l}^{\text{tot}} $
&$0.471108(1)$ & $0.178727(1)$ & $0.006398(1)$ \\ \hline\hline 
\end{tabular}
\caption{\label{testPHOTON} Decay widths in the case of photon that can be tested.}
\end{table}

\begin{figure}[hbtp]
\begin{center}$
\begin{array}{lr}
\dfrac{d\Gamma}{dE_{\gamma}} & \\
\includegraphics[width=15cm, height=6cm]
{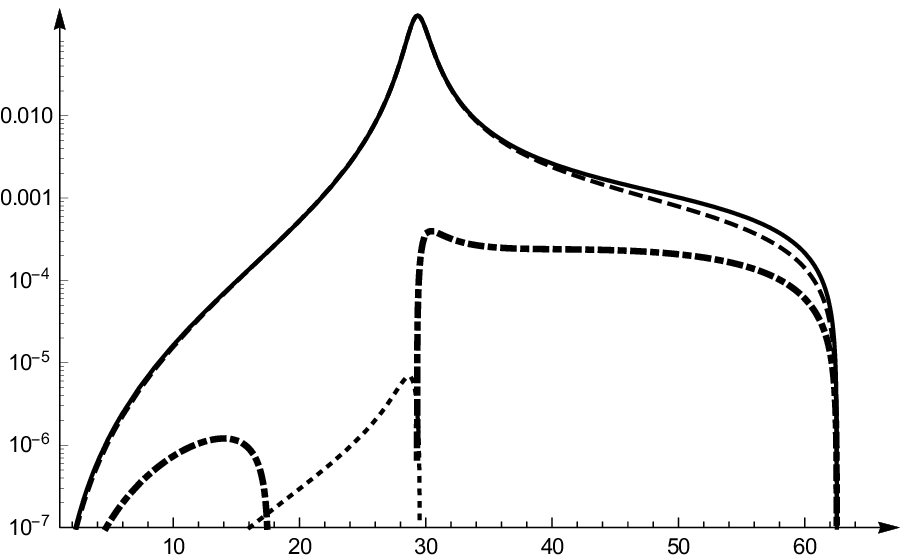}\\
&\hspace{-1cm}E_{\gamma} [GeV]
\end{array}$
\end{center}
\caption{\label{diffGam} Differential decay width as a function of 
$E_{\gamma}$ }
\end{figure}

In this Fig.~\ref{diffGam}, differential decay widths 
in respective to $E_{\gamma}$ are presented. The solid line 
denotes for contributing of all diagrams. The dashed line
shows for the contributions of triangle diagrams. While 
the dot-dashed line presents for the interference between 
three-point diagrams and box diagrams in Fig.~\ref{boxW}. 
The dotted line is for the amplitude of the interference between 
three-point diagrams and box diagrams in Fig.~\ref{boxF}.  
We find that the dominant contributions of 
triangle diagrams. We observe the peak 
which is corresponding to the $Z$-pole
that $Z$ boson decay to neutrinos. The position 
of this peak is at 
$E_{\gamma}= \frac{M_H^2-M_Z^2}{2M_H}=29.3159$ GeV.
Beyond this peak, the contributions of the interference 
between three-point diagrams and box diagrams are visible. 
We note that all results present here are for a family of neutrino 
in final state. For all neutrinos, we simply multiply factor 
$3$ for all above results. The physical results are important
for testing standard model at modern energy and evaluating 
precisely for the standard model background.

Furthermore, it is stress that general
results for one-loop contributing to the mentioned decay 
processes in arbitrary beyond the standard model have 
presented in~\cite{Phan:2021xwc}.
\section{Conclusions}   
One-loop contributions to the decay 
$H\rightarrow \nu_l\bar{\nu}_l\gamma$ with $l=e, \mu, \tau$
within standard model framework have revisited in this paper. 
Two representations for the form factors have computed in 
this calculation. As a result, the calculations are not 
only checked numerically by verifying the ultraviolet 
finiteness of the results but also confirming the ward 
identity of the amplitude. We find that the results satisfy
the ward identity and are good stability with varying ultraviolet 
cutoff parameters. In phenomenological results, all the 
physical results have examined with the present input 
parameters. In detail, we have studied the partial decay 
widths for the decay channels in both cases of the detected 
photon and invisible photon. Differential decay widths 
have also generated as a function of
energy of final photon. The physical results are important
for testing standard model at modern energy and evaluating 
precisely for the standard model background. We have also 
discussed that one applies hypergeometric functions  
for scalar one-loop integrals at general space-time 
dimensions $d$, the results in this paper can be 
perfromed higher-power of $\varepsilon$-expansion 
that may be taken into account in two-loop and 
higher-loop contributing 
to the mentioned channels. 
\\

\noindent
{\bf Acknowledgment:}~
This research is funded by Vietnam National Foundation
for Science and Technology Development (NAFOSTED) under 
the grant number $103.01$-$2019.346$.\\
\appendix 
\section{Reduction for scalar PV-functions}  %
In this appendix, we present all reduction formulas 
for scalar PV-functions expressing in terms of 
scalars one-loop integrals $A_0, B_0, C_0$ and $D_0$. 
We first show reduction for all $A, B$-functions:
\begin{eqnarray}
A_{00}(M_W^2)&=& \dfrac{M_W^2 }{d}A_0(M_W^2), \\
\n\\
B_1(M_H^2,M_W^2,M_W^2) &=&
-\dfrac{1}{2} B_0(M_H^2,M_W^2,M_W^2),        
\\
\n\\
B_{00}(p^2 ,M_W^2,M_W^2) 
&=&\dfrac{1}{4 (d-1)}
\Big\{ 
2 A_0(M_W^2)-(p^2-4 M_W^2) B_0(p^2,M_W^2,M_W^2)
\Big\}, \\
\n\\
B_{11}(M_H^2,M_W^2,M_W^2) &=&
\\
&&
\hspace{-3cm}=
\dfrac{1}{4 M_H^2 (d-1)}
\Big\{
(d M_H^2-4 M_W^2) B_0(M_H^2,M_W^2,M_W^2) 
+ 2 (d-2) A_0(M_W^2) 
\Big\}, 
\n \\
\n\\
B_{001}(M_H^2,M_W^2,M_W^2)
&=&
\dfrac{1}{8 (d-1)}
\Big\{ 
(M_H^2-4 M_W^2) B_0(M_H^2,M_W^2,M_W^2)-2 A_0(M_W^2)
\Big\},  \\
\n\\
B_{111}(M_H^2,M_W^2,M_W^2) &=&
\n  \\
&&\hspace{-4cm}
=
\dfrac{1}{8 M_H^2 (1-d)} \Big\{
\Big[(d+2) M_H^2-12 M_W^2 \Big] 
B_0(M_H^2,M_W^2,M_W^2)   
+6 (d-2) A_0(M_W^2)  \Big\}. 
\end{eqnarray}
Here $p^2 =M_H^2, 0$. We next have the following 
relations for 
$C$-functions:
\begin{eqnarray}
C_1(0,q_{13},M_H^2,M_W^2,0,M_W^2) 
&=&
\dfrac{1}{(M_H^2-q_{13})^2}
\Big[
(M_H^2-q_{13}) B_0(0,M_W^2,0)
 \\
&&
+(M_H^2+q_{13}) B_0(q_{13},0,M_W^2)
-2 M_H^2 B_0(M_H^2,M_W^2,M_W^2)
\n \\
&&
+M_H^2 (2 M_W^2-M_H^2+q_{13}) 
C_0(0,q_{13},M_H^2,M_W^2,0,M_W^2)
\Big], \n 
\\
\n\\
C_2(0,q_{13},M_H^2,M_W^2,0,M_W^2) 
&=&\dfrac{1}{M_H^2-q_{13}} \Big[
B_0(M_H^2,M_W^2,M_W^2)
-B_0(q_{13},0,M_W^2)
\n \\
&& 
-M_W^2 C_0(0,q_{13},M_H^2,M_W^2,0,M_W^2)
\Big], \\
\n\\
C_{2}(q_{12},0,M_H^2,M_W^2,M_W^2,M_W^2)
&=&
\dfrac{1}{(M_H^2-q_{12})^2}
\Big\{
(q_{12}-M_H^2) B_0(0,M_W^2,M_W^2)
\n \\
&&
+(M_H^2+q_{12}) B_0(M_H^2,M_W^2,M_W^2)
-2 q_{12} B_0(q_{12},M_W^2,M_W^2)
\n \\
&&
+q_{12} (M_H^2-q_{12}) 
C_0(q_{12},0,M_H^2,M_W^2,M_W^2,M_W^2)
\Big\},\n \\
\n\\
C_{00}(0,q_{12},M_H^2,M^2,M^2,M^2)
&=&
\dfrac{1}{2 (d-2) (M_H^2-q_{12})}
\times
\n \\
&&\times
\Big\{
M_H^2 B_0(M_H^2,M^2,M^2)-q_{12} B_0(q_{12},M^2,M^2)
\\
&&\hspace{0cm} 
+2 M^2 (M_H^2-q_{12}) C_0(0,q_{12},M_H^2,M^2,M^2,M^2)
\Big\}, \n \\
\n\\
C_{00}(q_{12},0,M_H^2;M^2,M^2,M^2)
&=& C_{00}(0,q_{12},M_H^2;M^2,M^2,M^2), 
\\
\n\\
C_{12}(0,q_{12},M_H^2,M^2,M^2,M^2)
&=&
\dfrac{1}{2 (d-2) (M_H^2-q_{12})^2}
\Big\{
-2 M_H^2 B_0(M_H^2,M^2,M^2)
\n \\
&& + \Big[(d-2) M_H^2-(d-4) q_{12} \Big] B_0(q_{12},M^2,M^2)
\\
&&
+4 M^2 (q_{12}-M_H^2) C_0(0,q_{12},M_H^2,M^2,M^2,M^2)
\Big\},
\n \\
\n\\
C_{22}(0,q_{12},M_H^2,M^2,M^2,M^2)
&=&
\dfrac{1}{2 (M_H^2-q_{12})}
\Big\{
B_0(q_{12},M^2,M^2)-B_0(M_H^2,M^2,M^2)
\Big\}. \n\\
\end{eqnarray}
Where $M^2=M_W^2, m_f^2$. Further, we also 
have 
\begin{eqnarray}
C_{12}(q_{12},0,M_H^2,M_W^2,M_W^2,M_W^2)
&=&
\dfrac{1}{(d-2) (M_H^2-q_{12})^3}
\Big\{
(d-2) (M_H^2-q_{12}) A_0(M_W)
\n \\
&&
+(d-2) (M_H^2+M_W^2) 
(M_H^2-q_{12}) B_0(0,M_W^2,M_W^2)
\n \\
&&
+\Big[
(3-2 d) M_H^2 q_{12} 
-M_H^4 \Big]
B_0(M_H^2,M_W^2,M_W^2)
 \\
&&
+\Big[
(2 d -3) M_H^2 q_{12} 
+q_{12}^2 \Big]
B_0(q_{12},M_W^2,M_W^2)
\n \\
&&
+(q_{12} - M_H^2) \Big[2 M_H^2 M_W^2+ M_H^2 q_{12} 
(d-2) +2 M_W^2 q_{12}\Big]
\n \\
&& 
\times C_0(q_{12},0,M_H^2,M_W^2,M_W^2,M_W^2)
\Big\}, \n \\
\n\\
C_{22}(q_{12},0,M_H^2,M_W^2,M_W^2,M_W^2)
&=&
\dfrac{1}{2 (d-2) (M_H^2-q_{12})^3}
\Big\{
(4-2d) (M_H^2-q_{12}) A_0(M_W^2)
\n \\
&& 
+ (4-2d) (M_H^2-q_{12}) 
(M_W^2+q_{12}) B_0(0,M_W^2,M_W^2)
\n \\
&&
+ 4 q_{12}^2 (1-d) 
B_0(q_{12},M_W^2,M_W^2)
\n \\
&&
+\Big[
(2-d) (M_H^4 - 3 q_{12}^2) 
+2 d M_H^2 q_{12} 
\Big]
B_0(M_H^2,M_W^2,M_W^2)
\n \\
&&
+2 q_{12} (M_H^2-q_{12}) 
\Big[4 M_W^2+(d-2) q_{12}\Big]
\times \n 
\\
&&
\times 
C_0(q_{12},0,M_H^2,M_W^2,M_W^2,M_W^2)
\Big\}. \n
\end{eqnarray}
We finally devote for all relations for 
$D$-functions. In this paper, we consider
reduction formulas for all $D$-functions 
involving the decay processes in the limit 
of $m_l\rightarrow 0$. In detail, the relations 
are shown as follows:
\begin{eqnarray}
D_{1}(0,0,0,M_H^2;q_{12},q_{13};M_W^2,0,M_W^2,M_W^2)
&=&
\dfrac{1}{2 q_{12} q_{13} q_{23}}
\times
\n \\
&&\hspace{-6.5cm} \times
\Big\{
q_{13} (M_H^2-q_{12}) C_0(0,0,q_{13};0,M_W^2,M_W^2)
+q_{12} (q_{13}- q_{23}) C_0(0,0,q_{12},M_W^2,0,M_W^2)
\n \\
&&\hspace{-6.5cm}
+\Big[
q_{12} (q_{12}+q_{23})
-M_H^2 (q_{12}-q_{23})
\Big] C_0(0,q_{13},M_H^2,M_W^2,0,M_W^2)
\n \\
&&\hspace{-6.5cm}
-(M_H^2-q_{12})^2 C_0(q_{12},0,M_H^2,M_W^2,M_W^2,M_W^2)
\\
&&\hspace{-6.5cm}
+(M_H^2-q_{12}) \Big[
M_H^2 (M_W^2-q_{12})+q_{12} (-M_W^2+q_{12}+q_{23})
\Big]
\times
\n \\
&&\hspace{-0.5cm} 
\times D_0(0,0,0,M_H^2;q_{12},q_{13}; M_W^2,0,M_W^2,M_W^2)
\Big\}, 
\n \\
\n \\
D_{1}(0,0,0,M_H^2;q_{23},q_{13};M_W^2,0,0,M_W^2)
&=&
\dfrac{1}{2 q_{12} q_{13} q_{23}}
\times
\n \\
&&\hspace{-6.5cm} \times
\Bigg\{
q_{13} (q_{12}+q_{13}) C_0(0,0,q_{13},0,0,M_W^2)
+q_{23} (q_{13} - q_{12}) C_0(0,0,q_{23},M_W^2,0,0)
\n \\
&&\hspace{-6.5cm}
+
(M_H^2 q_{12}- q_{13} q_{23}) C_0(0,q_{13},M_H^2,M_W^2,0,M_W^2)
\n \\
&&\hspace{-6.5cm}
-(q_{12}+q_{13})^2 C_0(q_{23},0,M_H^2,M_W^2,0,M_W^2)
\n \\
&&\hspace{-6.5cm}
-\Big\{
M_H^4 (q_{23}-M_W^2)
+M_H^2 \Big[M_W^2 (q_{12}+q_{23})-q_{23} (q_{12}+2 q_{23})\Big]
\\
&&\hspace{-6.5cm}
+q_{23} \Big[M_W^2 q_{12}+q_{23} (q_{12}+q_{23})\Big]
\Big\}
 D_0(0,0,0,M_H^2;q_{23},q_{13}; M_W^2,0,0,M_W^2)
\Bigg\},
\n \\
\n \\
D_{2}(0,0,0,M_H^2;q_{12},q_{13};M_W^2,0,M_W^2,M_W^2) 
&=&
\dfrac{1}{2 q_{12} q_{13} q_{23}}
\times
\n \\
&&\hspace{-6.5cm} \times
\Big\{
q_{13} (q_{12}-q_{23}) C_0(0,0,q_{13};0,M_W^2,M_W^2)
- (q_{12}+q_{23})^2 C_0(0,q_{13},M_H^2,M_W^2,0,M_W^2)
\n \\
&&\hspace{-6.5cm}
+\Big[
q_{12} (q_{12}+q_{23})
-M_H^2 (q_{12}-q_{23})
\Big] C_0(q_{12},0,M_H^2,M_W^2,M_W^2,M_W^2)
\n \\
&&\hspace{-6.5cm}
+q_{12} (q_{12}+q_{23}) C_0(0,0,q_{12},M_W^2,0,M_W^2)
\n \\
&&\hspace{-6.5cm}
+\Big[
M_H^2 M_W^2 (q_{12}-q_{23})- M_H^2 q_{12} (q_{12}+q_{23})
+q_{12} (q_{12}+q_{23}) 
(-M_W^2+q_{12}+q_{23})
\Big] \times
\n \\
&&\hspace{-6.5cm}
\times
D_0(0,0,0,M_H^2; q_{12},q_{13}; 
M_W^2,0,M_W^2,M_W^2)
\Big\},
\n \\
\n \\
D_{2}(0,0,0,M_H^2;q_{23},q_{13};M_W^2,0,0,M_W^2)
&=&
\dfrac{1}{2 q_{12} q_{13} q_{23} }
\times  \\
&&\hspace{-6.5cm} \times
\Bigg\{
-(q_{12}+q_{23})^2 C_0(0,q_{13},M_H^2,M_W^2,0,M_W^2)
- q_{13}(q_{12}-q_{23}) C_0(0,0,q_{13},0,0,M_W^2)
\n \\
&&\hspace{-6.5cm}
+(M_H^2 q_{12}- q_{13} q_{23}) C_0(q_{23},0,M_H^2,M_W^2,0,M_W^2)
+q_{23} (q_{12}+q_{23}) C_0(0,0,q_{23},M_W^2,0,0)
\n \\
&&\hspace{-6.5cm}
+\Big\{(q_{12}+q_{23}) \Big[M_W^2 q_{12}+q_{23} (q_{12}+q_{23})\Big]
-M_H^2 \Big[M_W^2 (q_{12}-q_{23})+q_{23} (q_{12}+q_{23})\Big]\Big\}
\times
\n \\
&&\hspace{-6.5cm}
\times 
D_0(0,0,0,M_H^2;q_{23},q_{13}; M_W^2,0,0,M_W^2)
\Bigg\}, 
\n \\
\n \\
D_{2}(0,0,0,M_H^2;q_{23},q_{12};M_W^2,M_W^2,0,M_W^2) 
&=&
-\dfrac{1}{2 q_{12} q_{13} q_{23} } \times
\n \\
&&\hspace{-6.5cm} \times
\Bigg\{
q_{12} (q_{13}-q_{23}) C_0(0,0,q_{12},M_W^2,0,M_W^2)
+q_{23} (q_{12}-M_H^2) C_0(0,0,q_{23},M_W^2,M_W^2,0)
\n \\
&&\hspace{-6.5cm} 
+(M_H^2-q_{12})^2 C_0(0,q_{12},M_H^2,M_W^2,M_W^2,M_W^2)
\n \\
&&\hspace{-6.5cm} 
+(q_{12} q_{23}-M_H^2 q_{13})
C_0(q_{23},0,M_H^2,M_W^2,0,M_W^2)  \\
&&\hspace{-6.5cm} 
+(q_{12}-M_H^2) \Big[M_H^2 M_W^2-q_{12} (M_W^2+q_{23})\Big] 
D_0(0,0,0,M_H^2; q_{23},q_{12}; M_W^2,M_W^2,0,M_W^2)
\Bigg\},
\n \\
D_{3}(0,0,0,M_H^2;q_{12},q_{13};M_W^2,0,M_W^2,M_W^2)
&=&
\dfrac{1}{2 q_{13} q_{23}}
\times
\n \\
&&\hspace{-6.5cm} \times
\Big\{
(q_{12}+q_{23}) C_0(0,q_{13},M_H^2,M_W^2,0,M_W^2)
- q_{13} C_0(0,0,q_{13},0,M_W^2,M_W^2)
\n \\
&&\hspace{-6.5cm}
+(q_{13}-q_{23}) C_0(q_{12},0,M_H^2,M_W^2,M_W^2,M_W^2)
-q_{12} C_0(0,0,q_{12},M_W^2,0,M_W^2)
\\
&&\hspace{-6.5cm}
+\Big[
M_W^2 (q_{12}+2 q_{23})
- M_H^2 (M_W^2-q_{12})
-2q_{12} (q_{12}+q_{23})
\Big]
\times
\n \\
&&\hspace{-6.5cm} 
\times
D_0(0,0,0,M_H^2; q_{12},q_{13}; M_W^2,0,M_W^2,M_W^2)
\Big\}, 
\n \\
\n \\
D_{3}(0,0,0,M_H^2;q_{23},q_{12};M_W^2,M_W^2,0,M_W^2) 
&=& -\dfrac{1}{2 q_{12} q_{13}} \times
\n \\
&&\hspace{-6.5cm} \times
\Bigg\{
(q_{13} - q_{12}) C_0(q_{23},0,M_H^2,M_W^2,0,M_W^2)
+q_{12} C_0(0,0,q_{12},M_W^2,0,M_W^2)
\\
&&\hspace{-6.5cm} 
+q_{23} C_0(0,0,q_{23},M_W^2,M_W^2,0)
-(M_H^2-q_{12}) C_0(0,q_{12},M_H^2,M_W^2,M_W^2,M_W^2)
\n \\
&&\hspace{-6.5cm} 
+\Big[M_W^2 (M_H^2-q_{12}) 
-q_{12} q_{23}\Big]
D_0(0,0,0,M_H^2;q_{23},q_{12}; M_W^2,M_W^2,0,M_W^2)
\Bigg\}, 
\n \\
\n \\
D_{3}(0,0,0,M_H^2;q_{23},q_{13};M_W^2,0,0,M_W^2) 
&=&
-\dfrac{1}{2 q_{12} q_{13}}
\times
\n \\
&&\hspace{-6.5cm} \times
\Big\{
q_{13} C_0(0,0,q_{13},0,0,M_W^2)
-(q_{12}+q_{23}) C_0(0,q_{13},M_H^2,M_W^2,0,M_W^2)
\n \\
&&\hspace{-6.5cm}
+ (q_{12}-q_{13}) C_0(q_{23},0,M_H^2,M_W^2,0,M_W^2)
+q_{23} C_0(0,0,q_{23},M_W^2,0,0)
\\
&&\hspace{-6.5cm}
+\Big[ M_W^2 (q_{13}+q_{23})-q_{23} q_{13}\Big] 
D_0(0,0,0,M_H^2; q_{23},q_{13}; M_W^2,0,0,M_W^2)
\Big\}.
\n
\end{eqnarray}
For one-loop tensor four-point with rank $P=2$, we have
\begin{eqnarray}
D_{00}(0,0,0,M_H^2;q_{12},q_{13};M_W^2,0,M_W^2,M_W^2) 
&=&
\dfrac{1}{4 (3-d) q_{12} q_{13} q_{23}}
\times
\n \\
&&\hspace{-6cm} \times
\Bigg\{ 
\Big[M_W^2 q_{12} (q_{13}-q_{23})- M_W^2 q_{23} (q_{13}+q_{23}) 
- q_{12} q_{13} (q_{12}+q_{23})   \Big] 
\n\\
&& \hspace{-6cm} \times 
C_0(0,q_{13},M_H^2,M_W^2,0,M_W^2)
\n \\
&&\hspace{-6cm} 
+(q_{13}+q_{23}) 
\Big[M_W^2 (q_{13}+q_{23})-q_{12} q_{13}\Big] 
C_0(q_{12},0,M_H^2,M_W^2,M_W^2,M_W^2)
\n \\
&&\hspace{-6cm} 
+ q_{12} \Big[M_W^2 (q_{23}-q_{13})+q_{12} q_{13}\Big] 
C_0(0,0,q_{12},M_W^2,0,M_W^2) \\
&&\hspace{-6cm} 
+q_{13} \Big[q_{12} q_{13}-M_W^2 (q_{13}+q_{23})\Big] 
C_0(0,0,q_{13},0,M_W^2,M_W^2)
\n \\
&&\hspace{-6cm} 
-\Big[M_W^4 (q_{13}+q_{23})^2 
+ 2 M_W^2 q_{12} q_{13} (q_{23}-q_{13})+q_{12}^2 q_{13}^2\Big]
\times
\n \\
&&\hspace{-1cm} 
\times D_0(0,0,0,M_H^2;q_{12},q_{13}; M_W^2,0,M_W^2,M_W^2)
\Bigg\},  \n \\
\n\\
D_{00}(0,0,0,M_H^2;q_{23},q_{12};M_W^2,M_W^2,0,M_W^2) 
&=&
\dfrac{1}{4 (3-d) q_{12} q_{13} q_{23}}
\times
\n \\
&&\hspace{-6cm} \times
\Big\{
(q_{13}+q_{23}) \Big[M_W^2 (q_{13}+q_{23})-q_{12} q_{23}\Big] 
C_0(0,q_{12},M_H^2,M_W^2,M_W^2,M_W^2)
\n \\
&&\hspace{-6cm} 
-\Big[M_W^2 q_{12} (q_{13}-q_{23})+ M_W^2 q_{13}
(q_{13}+q_{23})+q_{12} q_{23} (q_{12}+q_{13})
\Big]
\n\\
&& \hspace{-6cm} \times 
C_0(q_{23},0,M_H^2,M_W^2,0,M_W^2)
\n \\
&&\hspace{-6cm} 
+q_{12} \Big[M_W^2 (q_{13}-q_{23})+q_{12} q_{23}\Big] 
C_0(0,0,q_{12},M_W^2,0,M_W^2)
\\
&&\hspace{-6cm} 
+q_{23} \Big[q_{12} q_{23}-M_W^2 (q_{13}+q_{23})\Big] 
C_0(0,0,q_{23},M_W^2,M_W^2,0)
\n \\
&&\hspace{-6.0cm} 
-\Big[M_W^4 (q_{13}+q_{23})^2+2 M_W^2 q_{12} q_{23} 
(q_{13}-q_{23})+q_{12}^2 q_{23}^2\Big]
\times
\n \\
&&\hspace{-1cm} \times
D_0(0,0,0,M_H^2;q_{23},q_{12}; M_W^2,M_W^2,0,M_W^2)
\Big\},  
\n \\
\n\\
D_{00}(0,0,0,M_H^2;q_{23},q_{13};M_W^2,0,0,M_W^2) 
&=&
\dfrac{1}{4 (3-d) q_{12} q_{13} q_{23}}
\times
\n \\
&&\hspace{-6cm} \times
\Bigg\{
\Big[ M_W^2  q_{12} (q_{23}-q_{13}) 
+ M_W^2  q_{23} (q_{13}+q_{23})
-q_{13} q_{23} (q_{12}+q_{23})
\Big]
\times 
\n \\
&& 
\hspace{-6cm}
\times 
C_0(0,q_{13},M_H^2,M_W^2,0,M_W^2)
\n \\
&&\hspace{-6cm} 
+\Big\{
M_W^2 \Big[q_{12} (q_{13}-q_{23})
+q_{13} (q_{13}+q_{23})\Big] 
-q_{13} q_{23} (q_{12}+q_{13})
\Big\}
\times 
\n \\
&& 
\hspace{-6cm}
\times 
C_0(q_{23},0,M_H^2,M_W^2,0,M_W^2)
\n \\
&&\hspace{-6cm} 
+q_{13} \Big[q_{13} q_{23}-M_W^2 (q_{13} 
+ q_{23})\Big] C_0(0,0,q_{13},0,0,M_W^2)
\\
&&\hspace{-6cm} 
+q_{23} \Big[q_{13} q_{23} 
- M_W^2 (q_{13}+q_{23})\Big] C_0(0,0,q_{23},M_W^2,0,0)
\n \\
&&\hspace{-6cm} 
-\Big[q_{13} q_{23}-M_W^2 (q_{13} 
+ q_{23})\Big]^2 D_0(0,0,0,M_H^2;q_{23},q_{13}; M_W^2,0,0,M_W^2)
\Big\},
\n \\
\n\\
D_{12}(0,0,0,M_H^2;q_{23},q_{13};M_W^2,0,0,M_W^2)
&=&
\dfrac{M_H^2 }{q_{12} q_{13} q_{23}} B_0(M_H^2,M_W^2,M_W^2)
\n \\
&&\hspace{-6cm} 
-\dfrac{(q_{12}+q_{13}) }{q_{12} q_{13} q_{23}} B_0(q_{13},0,M_W^2) 
-\dfrac{(q_{12}+q_{23})} {q_{12} q_{13} q_{23}}
 B_0(q_{23},M_W^2,0)
\n \\
&&\hspace{-6cm} 
+
\dfrac{1}{4 (3-d) q_{12}^2 q_{13} q_{23}^2}
\Big\{
M_W^2 q_{12}^2 
\Big[(d-2) q_{13}+(10-3 d) q_{23}\Big]
\n\\
&&\hspace{-6cm}
+M_W^2 q_{12} 
\Big[ (d-2) (q_{13}^2
+ q_{23}^2)
-2 q_{13} q_{23} (d-4)
\Big]
-(d-2) M_W^2 q_{13} q_{23} (q_{13}+q_{23})
\n \\
&&\hspace{-6cm} 
+q_{13} q_{23} 
\Big[
(d-4) M_H^2 q_{12}
+(d-2) q_{13} q_{23}
\Big]
\Big\} 
C_0(0,0,q_{13},0,0,M_W^2)
\n \\
&&
\hspace{-6cm}  
+
\dfrac{1}
{4 (3-d) q_{12}^2 q_{13}^2 q_{23}}
\Big\{
M_W^2 q_{12}^2 
\Big[(d-2) q_{23}+(10-3 d) q_{13}\Big]
\n \\
&&\hspace{-6cm} 
+M_W^2 q_{12}
\Big[
(d-2) (q_{13}^2
+ q_{23}^2)
-2 q_{13} q_{23} (d-4)
\Big]
-(d-2) M_W^2 
q_{13} q_{23} (q_{13}+q_{23})
\n \\
&&\hspace{-6cm} 
+q_{13} q_{23} 
\Big[
(d-4) M_H^2 q_{12}
+(d-2) q_{13} q_{23}
\Big]
\Big\}  
C_0(0,0,q_{23},M_W^2,0,0)
\n \\
&&\hspace{-6cm} 
+
\dfrac{1}
{4 (d-3) q_{12}^2 q_{13}^2 q_{23}^2}
\Big\{
q_{13} q_{23} (q_{12}+q_{23}) 
\Big[(d-4) M_H^2 q_{12}+(d-2) q_{13} q_{23}\Big]
\n\\
&&
\hspace{-6cm}
-M_W^2 q_{12}^3 (q_{13}-q_{23}) (d-2)
-M_W^2 q_{12}^2 
\Big[
(d-2) (q_{13}^2
-2 q_{23}^2)
+(3 d-10) q_{13} q_{23}
\Big]
\n\\
&&
\hspace{-6cm}
-M_W^2 q_{12} q_{23} 
\Big[
2 q_{13}^2 (d-4)
+(3 d-10) q_{13} q_{23}
-(d-2) q_{23}^2
\Big]
\n \\
&&
\hspace{-6cm}
- M_W^2 q_{13} q_{23}^2 (q_{13}+q_{23}) (d-2)
\Big\}
C_0(0,q_{13},M_H^2,M_W^2,0,M_W^2)
\n \\
&&\hspace{-6cm} 
+
\dfrac{1}{4 (d-3) q_{12}^2 q_{13}^2 q_{23}^2}
\Big\{
(d-2) M_W^2 q_{12} q_{13} (q_{12}+q_{13})^2
\n \\
&&\hspace{-6cm} 
- M_H^2 M_W^2 q_{23}
\Big[
(d-2)(q_{12}^2
+ q_{13}^2)
+2 q_{12} q_{13} (d-4)
\Big]
\n \\
&&\hspace{-6cm}
+q_{13} q_{23} (q_{12}+q_{13}) 
\Big[
(d-4) M_H^2 q_{12}+(d-2) q_{13} q_{23}
\Big]
\Big\}
C_0(q_{23},0,M_H^2,M_W^2,0,M_W^2)
\n \\
&&\hspace{-6cm} 
+
\dfrac{1}{4 (d-3) q_{12}^2 q_{13}^2 q_{23}^2}
\Big\{
(d-2) M_W^4 q_{13} q_{23} (q_{13}+q_{23})^2
\n \\
&&\hspace{-6cm}
- M_H^2 M_W^4 q_{12}
\Big[
(d-2) (q_{13}^2+q_{23}^2)-2 q_{13} q_{23} (d-4)
\Big]
\n \\
&&\hspace{-6cm}
+2 M_W^2 q_{13} q_{23} 
\Big[q_{12}^2 (q_{13}+q_{23})
+q_{12} \Big(
q_{13}^2-2 (d-4) q_{13} q_{23}+q_{23}^2
\Big)
\n \\
&&\hspace{-6cm} 
-(d-2) q_{13} q_{23} (q_{13}+q_{23})\Big] 
+q_{13}^2 q_{23}^2 
\Big[(d-4) M_H^2 q_{12}+(d-2) q_{13} q_{23}\Big]
\Big\} \times
\n \\
&&\hspace{-6cm} \times
D_0(0,0,0,M_H^2;q_{23},q_{13}; M_W^2,0,0,M_W^2),
\n \\
\n \\
D_{12}(0,0,0,M_H^2;q_{23},q_{12};M_W^2,M_W^2,0,M_W^2) 
&=&
\n \\
&&\hspace{-6cm} 
=\dfrac{1}{q_{12} q_{23}} B_0(0,M_W^2,0)
-\dfrac{(q_{13}+q_{23}) }
{q_{12} q_{13} q_{23}} B_0(q_{23},M_W^2,0)
\n \\
&&\hspace{-6cm} +
\dfrac{M_H^2 }{q_{12} q_{13} q_{23}}
B_0(M_H^2,M_W^2,M_W^2)
-\dfrac{(q_{12}+q_{13}) }{q_{12} q_{13} q_{23}}
B_0(q_{12},M_W^2,M_W^2)
\n \\
&&\hspace{-6cm}
+
\dfrac{1}{4 (d-3) q_{12} q_{13}^2 q_{23}^2}
\Big\{
q_{12} M_W^2
\Big[
(d-2) (q_{13}^2
+ q_{23}^2)
+2 q_{13} q_{23} (d-4)
\Big]
\n \\
&&\hspace{-6cm}
-q^2_{12} q_{23} 
\Big[(d-4) 
q_{13}+(d-2) q_{23}\Big]
-(d-4) q_{12} q_{23} q_{13} (q_{13}+q_{23})
\n \\
&&\hspace{-6cm}
+ q_{13} M_W^2 
\Big[
(d-2) (q_{13}^2
- q_{23}^2)
+4 q_{13} q_{23} (d-3)
\Big] 
\Big\}  
C_0(0,0,q_{12},M_W^2,0,M_W^2)
\n \\
&&\hspace{-6cm} 
+
\dfrac{1}{4 (3-d) q_{12}^2 q_{13}^2 q_{23}}
\Big\{ (d-2) M_W^2 (q_{13}+q_{23}) 
\Big[q_{12} (q_{13}-q_{23})+q_{13} (q_{13}+q_{23})\Big]
\n \\
&&\hspace{-6.cm}
+q^2_{12} q_{23} 
\Big[(d-4) q_{13}+(d-2) q_{23}\Big]
+(d-4) q_{12} q_{13}  q_{23}(q_{13}+q_{23})
\Big\} \times
\n \\
&&\hspace{-6.cm} \times
C_0(0,0,q_{23},M_W^2,M_W^2,0)
\n \\
&&\hspace{-6cm} 
+
\dfrac{(q_{13}+q_{23})}
{4 (d-3) q_{12}^2 q_{13}^2 q_{23}^2} 
\Bigg\{
(d-2) M_W^2 (q_{13}+q_{23}) 
\Big[q_{12} (q_{13}-q_{23})+q_{13} (q_{13}+q_{23})\Big]
\n \\
&&\hspace{-6cm}
+q_{12}q_{23}  \Big[(d-4) q_{13}+(d-2) q_{23}\Big]
+(d-4) q_{13}q_{12} q_{23} (q_{13}+q_{23})
\Bigg\}
\n\\
&&\hspace{-6cm} 
\times
C_0(0,q_{12},M_H^2,M_W^2,M_W^2,M_W^2)
\n \\
&&\hspace{-6cm} +
\dfrac{1}
{4 (d-3) q_{12}^2 q_{13}^2 q_{23}^2}
\Bigg\{
q_{12} q_{23} (q_{12}+q_{13}) 
\Bigg[q_{12} \Big\{
(d-4) q_{13}+(d-2) q_{23}\Big]
\n\\
&&\hspace{-6cm}
+(d-4) q_{13} (q_{13}+q_{23})
\Big]
\\
&&\hspace{-6cm}
-M_W^2 \Big\{
q_{12}^2 
\Big[
(d-2) (q_{13}^2
+ q_{23}^2)
+2 (d-4) q_{13} q_{23}
\Big]
+(d-2) q_{13}^2 (q_{13}+q_{23})^2
\n \\
&&\hspace{-6cm}
+2 q_{12} q_{13} 
(q_{13}+q_{23}) \Big[(d-2) q_{13}+(d-4) q_{23}\Big]
\Big\}
\Bigg\}C_0(q_{23},0,M_H^2,M_W^2,0,M_W^2)
\n \\
&&\hspace{-6cm} 
+
\dfrac{1}
{4 (d-3) q_{12}^2 q_{13}^2 q_{23}^2}
\Bigg\{
(2-d) M_W^4 (q_{13}+q_{23})^2 
\Big[q_{12} (q_{13}-q_{23})+q_{13} 
(q_{13}+q_{23})\Big]
\n \\
&&\hspace{-6cm}
-2 M_W^2 q_{12} q_{23} 
\Big\{q_{12} 
\Big[q_{13}^2+(d-5) q_{13} q_{23} 
+ (d-2) q_{23}^2\Big]
+q_{13} (q_{13}^2- q_{23}^2)
\Big\}
\n \\
&&\hspace{-6cm}
+q_{12}^2 q_{23}^2 \Big\{
q_{12} \Big[(d-4) q_{13}+(d-2) q_{23}\Big]
+(d-4) q_{13} (q_{13}+q_{23})
\Big\}\; 
\Bigg\}\times 
\n \\
&&\hspace{-6cm}
\times 
D_0(0,0,0,M_H^2;q_{23},q_{12}; M_W^2,M_W^2,0,M_W^2),
\n \\
\n \\
D_{13}(0,0,0,M_H^2;q_{23},q_{13};M_W^2,0,0,M_W^2) &=&
-\dfrac{M_H^2}{q_{12} q_{13} (q_{12}+q_{23})}
B_0(M_H^2,M_W^2,M_W^2)
\n \\
&&\hspace{-6.5cm} 
+
\dfrac{1}{q_{12} (q_{12}+q_{23})}B_0(q_{13},0,M_W^2)
+
\dfrac{1}{q_{12} q_{13}}B_0(q_{23},M_W^2,0)
+
\n \\
&&\hspace{-6.5cm} 
\dfrac{1}{4 (d-3) q_{12}^2 q_{13} q_{23}}
\Big\{
q_{13} q_{23} \Big[(d-4) q_{12}+(d-2) q_{13}\Big]
\n \\
&&\hspace{-6.5cm}
-M_W^2 q_{12} \Big[(d-4) q_{13}-(d-2) q_{23}\Big]
-M_W^2 q_{13} (q_{13}+q_{23}) (d-2)
\Big\}
C_0(0,0,q_{13},0,0,M_W^2)
\n \\
&&\hspace{-6.5cm} 
+
\dfrac{1}{4 (d-3) q_{12}^2 q_{13}^2}
\Big\{
q_{13} q_{23} \Big[(d-4) q_{12}+(d-2) q_{13}\Big]
-M_W^2 q_{12} \Big[(d-4) q_{13}-(d-2) q_{23}\Big]
\n \\
&&\hspace{-6.5cm}
-M_W^2 q_{13} (q_{13}+q_{23}) (d-2)
\Big\}
C_0(0,0,q_{23},M_W^2,0,0)
\n \\
&&\hspace{-6.5cm} 
+
\dfrac{1}{4 (3-d) q_{12}^2 q_{13}^2 q_{23} (q_{12}+q_{23})}
\Big\{
M_W^2 q_{12}^3 \Big[(d-4) q_{13}+(d-2) q_{23}\Big]
\n \\
&&\hspace{-6.5cm} 
+(d-2) M_W^2 q_{12}^2 \Big[q_{13}^2-q_{13} q_{23}+2 q_{23}^2\Big]
-(d-2) M_W^2 q_{13} q_{23}^2 (q_{13}+q_{23})
\n \\
&&\hspace{-6.5cm} 
+M_W^2 q_{12} q_{23} \Big[4 q_{13}^2 (3-d)
+(8-3 d) q_{13} q_{23}+(d-2) q_{23}^2\Big]
\n \\
&&\hspace{-6.5cm} 
+q_{13} q_{23} (q_{12}+q_{23})^2 \Big[(d-4) q_{12}+(d-2) q_{13}\Big]
\Big\}C_0(0,q_{13},M_H^2,M_W^2,0,M_W^2)
\n \\
&&\hspace{-6.5cm} 
+ 
\dfrac{1}{4 (d-3) q_{12}^2 q_{13}^2 q_{23}}
\Big\{
M_W^2 q_{12}^2 \Big[(d-4) q_{13}+(d-2) q_{23}\Big]
\\
&&\hspace{-6.5cm}
+2 M_W^2 q_{12} q_{13} \Big[(d-3) q_{13}+(d-4) q_{23}\Big]
+(d-2) M_W^2 q_{13}^2 (q_{13}+q_{23})
\n \\
&&\hspace{-6.5cm}
-q_{13} q_{23} (q_{12}+q_{13}) \Big[(d-4) q_{12}+(d-2) q_{13}\Big]
\Big\}C_0(q_{23},0,M_H^2,M_W^2,0,M_W^2)
\n \\
&&\hspace{-6.5cm} 
+
\dfrac{q_{13} q_{23}-M_W^2 (q_{13}+q_{23})}
{4 (d-3) q_{12}^2 q_{13}^2 q_{23}}
\Big\{M_W^2 q_{12} \Big[(d-4) q_{13}-(d-2) q_{23}\Big]
\n \\
&&\hspace{-6.5cm} 
+ (d-2) M_W^2 q_{13} (q_{13}+q_{23})
-q_{13} q_{23} \Big[(d-4) q_{12}+(d-2) q_{13}\Big]
\Big\}
\n\\
&&\hspace{-6.5cm}
\times
D_0(0,0,0,M_H^2;q_{23},q_{13}; M_W^2,0,0,M_W^2),
\n \\
\n \\
D_{13}(0,0,0,M_H^2;q_{23},q_{12};M_W^2,M_W^2,0,M_W^2) 
&=&
-\dfrac{M_H^2}{q_{12} q_{13} (q_{13}+q_{23})}
B_0(M_H^2,M_W^2,M_W^2)
\n \\
&&\hspace{-6.5cm} 
+
\dfrac{1}{q_{13} (q_{13}+q_{23})}
B_0(q_{12},M_W^2,M_W^2)
+
\dfrac{1}{q_{12} q_{13}}
B_0(q_{23},M_W^2,0)
\n \\
&&\hspace{-6.5cm} 
+
\dfrac{1}{4 (d-3) q_{12} q_{13}^2 q_{23}}
\Big\{
q_{12} q_{23} \Big[(d-2) q_{12}+(d-4) q_{13}\Big]
-M_W^2 q_{12} \Big[(d-4) q_{13}+(d-2) q_{23}\Big]
\n \\
&&\hspace{-6.5cm} 
-(d-2) M_W^2 q_{13} (q_{13}-q_{23})
\Big\}
C_0(0,0,q_{12},M_W^2,0,M_W^2)
\n \\
&&\hspace{-6.5cm} 
+
\dfrac{1}
{4 (d-3) q_{12}^2 q_{13}^2}
\Big\{
M_W^2 q_{12} \Big[(d-4) q_{13}-(d-2) q_{23}\Big]
+(d-2) M_W^2 q_{13} (q_{13}+q_{23})
\n \\
&&\hspace{-6.5cm} 
+q_{12} q_{23} \Big[(d-2) q_{12}+(d-4) q_{13}\Big]
\Big\}C_0(0,0,q_{23},M_W^2,M_W^2,0)
\n \\
&&\hspace{-6.5cm} 
+
\dfrac{q_{13}+q_{23}}
{4 (3-d) q_{12}^2 q_{13}^2 q_{23}}
\Big\{
M_W^2 q_{12} \Big[(d-4) q_{13}-(d-2) q_{23}\Big]
+(d-2) M_W^2 q_{13} (q_{13}+q_{23})
\n \\
&&\hspace{-6.5cm} 
+q_{12} q_{23} \Big[(d-2) q_{12}+(d-4) q_{13}\Big]
\Big\} C_0(0,q_{12},M_H^2,M_W^2,M_W^2,M_W^2)
\\
&&\hspace{-6.5cm} +
\dfrac{1}{4 (d-3) q_{12}^2 q_{13}^2 q_{23}}
\Big\{
M_W^2 q_{12}^2 \Big[(d-4) q_{13}+(d-2) q_{23}\Big]
\n \\
&&\hspace{-6.5cm}
+2 M_W^2 q_{12} q_{13} \Big[(d-3) q_{13}+(d-4) q_{23}\Big]
-q_{12} q_{23} (q_{12}+q_{13}) \Big[(d-2) q_{12}+(d-4) q_{13}\Big]
\n \\
&&\hspace{-6.5cm}
+(d-2) M_W^2 q_{13}^2 (q_{13}+q_{23})
\Big\}C_0(q_{23},0,M_H^2,M_W^2,0,M_W^2)
\n \\
&&\hspace{-6.5cm} +
\dfrac{1}
{4 (d-3) q_{12}^2 q_{13}^2 q_{23}}, 
\Bigg\{
M_W^4 (q_{13}+q_{23}) \Big\{q_{12} \Big[(d-4) q_{13}-(d-2) q_{23}\Big]
\n \\
&&\hspace{-6.5cm} 
+(d-2) q_{13} (q_{13}+q_{23})\Big\}
-2 M_W^2 q_{12} q_{23} \Big\{
q_{12} \Big[q_{13}-(d-2) q_{23}\Big]
+q_{13} (q_{23}-q_{13})\Big\}
\n \\
&&\hspace{-6.5cm} 
-q_{12}^2 q_{23}^2 \Big[(d-2) q_{12}+(d-4) q_{13}\Big]
\Bigg\}D_0(0,0,0,M_H^2;q_{23},q_{12}; M_W^2,M_W^2,0,M_W^2),
\n \\
\n \\
D_{13}(0,0,0,M_H^2;q_{12},q_{13};M_W^2,0,M_W^2,M_W^2)
&=&
-\dfrac{M_H^2}{q_{13} q_{23}(q_{12} +q_{23})}
 B_0(M_H^2,M_W^2,M_W^2)
\n \\
&&\hspace{-6.5cm} 
 +
\dfrac{1}{q_{13} q_{23}}
B_0(q_{12},M_W^2,M_W^2) 
+
\dfrac{1}{q_{23}(q_{12} +q_{23})}
B_0(q_{13},0,M_W^2)
+
\n \\
&&\hspace{-6.5cm} 
\dfrac{1}{4 (d-3) q_{13}^2 q_{23}^2}
\Big\{
(d-2) \Big[ q_{13}^2 (q_{12}-M_W^2)
- M_W^2 q_{23}^2 \Big]
\n \\
&&\hspace{-6.5cm} 
+(d-4) q_{13} q_{23} (q_{12}-2 M_W^2)
\Big\}C_0(0,0,q_{12},M_W^2,0,M_W^2)
\n \\
&&\hspace{-6.5cm} +
\dfrac{1}{4 (d-3) q_{12} q_{13} q_{23}^2}
\Big\{
q_{12} q_{13} \Big[
(d-2) q_{13}+(d-4) q_{23}\Big]
\n \\
&&\hspace{-6.5cm} 
-(d-2) M_W^2 (q_{13}^2-q_{23}^2)
\Big\}C_0(0,0,q_{13},0,M_W^2,M_W^2)
\n \\
&&\hspace{-6.5cm} 
+
\dfrac{1}
{4 (d-3) q_{12} q_{13}^2 q_{23}^2 (q_{12}+q_{23})}
\Big\{
M_W^2 \Big[
q_{12}^2 [(d-2) 
(q_{13}^2+q_{23}^2)+2 q_{13} q_{23} (d-4)]
\\
&&\hspace{-6.5cm} 
+2 q_{12} q_{23} 
[2 q_{13}^2 (d-3)+(d-4) q_{13} q_{23}+(d-2) q_{23}^2]
+(d-2) q_{23}^2 (q_{23}^2-q_{13}^2)
\Big]
\n \\
&&\hspace{-6.5cm} 
-q_{12} q_{13} (q_{12}+q_{23})^2 \Big[
(d-2) q_{13}+(d-4) q_{23}
\Big]
\Big\}C_0(0,q_{13},M_H^2,M_W^2,0,M_W^2)
\n \\
&&\hspace{-6.5cm} +
\dfrac{1}
{4 (d-3) q_{12} q_{13}^2 q_{23}^2}
\Big\{
(q_{13}+q_{23}) \Big[
(d-2) M_W^2 (q_{13}^2-q_{23}^2)
\n \\
&&\hspace{-4.5cm}
-q_{12} q_{13} [(d-2) q_{13}+(d-4) q_{23}]
\Big]
\Big\}C_0(q_{12},0,M_H^2,M_W^2,M_W^2,M_W^2)
\n \\
&&\hspace{-6.5cm}
+
\dfrac{1}
{4 (3-d) q_{12} q_{13}^2 q_{23}^2}
\Big\{
(d-2) M_W^4 (q_{13}-q_{23}) (q_{13}+q_{23})^2
\n \\
&&\hspace{-6.5cm}
-2 M_W^2 q_{12} q_{13} \Big[
(d-2) q_{13}^2+(d-5) q_{13} q_{23}+q_{23}^2\Big]
\n \\
&&\hspace{-6.5cm}
+q_{12}^2 q_{13}^2 \Big[
(d-2) q_{13}+(d-4) q_{23}
\Big]
\Big\}D_0(0,0,0,M_H^2; q_{12},q_{13}; M_W^2,0,M_W^2,M_W^2),
\n \\
\n \\
D_{22}(0,0,0,M_H^2;q_{23},q_{12};M_W^2,M_W^2,0,M_W^2) 
&=&\n \\
&&\hspace{-6.5cm}
=
\dfrac{1}{q_{12} (q_{12}+q_{13})}
B_0(0,0,M_W^2)
-\dfrac{1}{q_{12} q_{23}}
B_0(0,M_W^2,0)
+
\dfrac{(q_{13}-q_{23})}{q_{12} q_{13} q_{23}}
 B_0(q_{12},M_W^2,M_W^2)
\n \\
&&\hspace{-6.5cm} 
- 
\dfrac{1}{q_{12} q_{13} q_{23} (q_{12}+q_{13})^2}
\times
\n\\
&&\hspace{-6.5cm} 
\times 
\Big\{q_{12}^2 (q_{13}-q_{23})
+q_{12} \Big[2 q_{13}^2+q_{13} q_{23}-q_{23}^2\Big]
+q_{13} (q_{13}+q_{23})^2\Big\}
B_0(M_H^2,M_W^2,M_W^2)
\n \\
&&\hspace{-6.5cm} +
\dfrac{1}{q_{12} q_{13} q_{23} 
(q_{12}+q_{13})^2}
\times
\n \\
&&\hspace{-6.5cm} 
\times
\Big\{q_{12}^2 q_{13}+q_{12} 
\Big[2 q_{13}^2+2 q_{13} q_{23} 
- q_{23}^2\Big]+q_{13} (q_{13}+q_{23})^2\Big\} 
B_0(q_{23},M_W^2,0)
\n \\
&&\hspace{-6.5cm} 
+
\dfrac{1}{4 (d-3) q_{12} q_{13}^2 q_{23}^2}
\Big\{q_{12} q_{23} 
\Big[
(3 d-10) q_{13}^2
-(d-2) q_{23} (2 q_{13}
- q_{23})
\Big]
\n \\
&&\hspace{-6.5cm}
-M_W^2 (q_{13}-q_{23}) 
\Big[
(d-2) (q_{13}^2
+ q_{23}^2)
+2 q_{13} q_{23} (3 d-8)
\Big]
\Big\}C_0(0,0,q_{12},M_W^2,0,M_W^2)
\n \\
&&\hspace{-6.5cm} +
\dfrac{(d-2) (q_{13}+q_{23})^2 }{4 (d-3) q_{12}^2 q_{13}^2 q_{23}}
\Big[M_W^2 (q_{13}+q_{23})-q_{12} q_{23}\Big]
\times \n \\
&&\hspace{-6.5cm} \times 
\Big[ 
C_0(0,0,q_{23},M_W^2,M_W^2,0)
-C_0(0,q_{12},M_H^2,M_W^2,M_W^2,M_W^2)\Big]
\n \\
&&\hspace{-6.5cm} +
\dfrac{1}
{4 (d-3) q_{12}^2 q_{13}^2 q_{23}^2 (q_{12}+q_{13})^2}
\times 
\n \\
&&\hspace{-6.5cm} \times
\Bigg\{
M_W^2 \Big[q_{12} (q_{13}-q_{23})+q_{13} (q_{13}+q_{23})\Big]
\Big\{q_{12}^2 
\Big[
(d-2) (q_{13}^2
+ q_{23}^2)
+2 q_{13} q_{23} (3 d-8)
\Big]
\n \\
&&\hspace{-6.5cm}
+2 q_{12} q_{13} (q_{13}+q_{23}) 
\Big[(d-2) q_{13}+(3 d-8) q_{23}\Big]
+(d-2) q_{13}^2 (q_{13}+q_{23})^2\Big\}
\n \\
&&\hspace{-6.5cm}
-q_{12} q_{23} (q_{12}+q_{13}) \Big\{q_{12}^2 
\Big[
(3 d-10) q_{13}^2
- (d-2) q_{23} (2 q_{13}
- q_{23})
\Big]
+(3 d-10) q_{13}^2 (q_{13}+q_{23})^2
\n \\
&&\hspace{-6.5cm}
+2 q_{12} q_{13} (q_{13}+q_{23}) \Big[
(3 d-10) q_{13}-(d-2) q_{23}\Big]
\Big\}
\Bigg\}\; C_0(q_{23},0,M_H^2,M_W^2,0,M_W^2)
\n \\
&&\hspace{-6.5cm} +
\dfrac{(q_{13}+q_{23})^2 }
{4 (d-3) q_{12}^2 q_{13}^2 q_{23}^2} 
 \Big\{
(d-2) \Big[M_W^4 (q_{13}+q_{23})^2
+ q_{12}^2 q_{23}^2\Big]
\n \\
&&\hspace{-6.5cm}
-2 M_W^2 q_{12} q_{23} \Big[(d-4) q_{13}+(d-2) q_{23}\Big]
\Big\} D_0(0,0,0,M_H^2;q_{23},q_{12}; M_W^2,M_W^2,0,M_W^2),
\n \\
\n \\
D_{22}(0,0,0,M_H^2;q_{23},q_{13};M_W^2,0,0,M_W^2)
&=&
\n \\
&&\hspace{-6.5cm}
=
\dfrac{1}{q_{13} (q_{12}+q_{13})}
B_0(0,0,M_W^2)
+
\dfrac{(q_{12}-q_{23})}{q_{12} q_{13} q_{23}}
B_0(q_{13},0,M_W^2) 
\n \\
&&\hspace{-6.5cm} 
+\dfrac{M_H^2 (q_{13} q_{23} - M_H^2 q_{12})}
{q_{12} q_{13} q_{23} (q_{12}+q_{13})^2}
B_0(M_H^2,M_W^2,M_W^2)
+
\dfrac{ (M_H^4 q_{12} 
-q_{13} q_{23}^2) }
{q_{12} q_{13} q_{23} (q_{12}+q_{13})^2}
B_0(q_{23},M_W^2,0)
\n \\
&&\hspace{-6.5cm} +
\dfrac{1}{4 (d-3) q_{12}^2 q_{13} q_{23}^2}
\Big\{
M_W^2 q_{12}^2 
\Big[(d-2) q_{13}+(10-3 d) q_{23}\Big]
\n \\
&&\hspace{-6.5cm}
+2 M_W^2 q_{12} q_{23} 
\Big[(d-2) q_{23}-(d-4) q_{13}\Big]
+ q_{13} q_{23} 
\Big[(3 d-10) q_{12}^2- (d-2) q_{23} (2 q_{12}+ q_{23})\Big]
\n \\
&&\hspace{-6.5cm} 
+(d-2) M_W^2 q_{23}^2 (q_{13}+q_{23})
\Big\} 
C_0(0,0,q_{13},0,0,M_W^2)
\n \\
&&\hspace{-6.5cm} 
+
\dfrac{(q_{12}+q_{23}) }{4 (d-3) q_{12}^2 q_{13}^2 q_{23}}
\Big\{
M_W^2 q_{12} \Big[(10-3 d) q_{13}+(d-2) q_{23}\Big]
\n \\
&&\hspace{-6.5cm}
+(d-2) q_{23} \Big[
M_W^2 (q_{13}+q_{23})
- q_{13} (q_{12}+q_{23})
\Big]
\Big\} C_0(0,0,q_{23},M_W^2,0,0)
\n \\
&&\hspace{-6.5cm} +
\dfrac{(d-2) (q_{12}+q_{23})^2}
{4 (d-3) q_{12}^2 q_{13}^2 q_{23}^2} 
\Big\{
M_W^2 \Big[q_{12} (q_{13}-q_{23})-q_{23} (q_{13}+q_{23})\Big]
\n \\
&&\hspace{-6.5cm}
+q_{13} q_{23} (q_{12}+q_{23})
\Big\}C_0(0,q_{13},M_H^2,M_W^2,0,M_W^2)
\n \\
&&\hspace{-6.5cm} +
\dfrac{1}
{4 (3-d) q_{12}^2 q_{13}^2 q_{23}^2 (q_{12}+q_{13})^2}
\times 
\n \\
&&\hspace{-6.5cm}
\Bigg\{M_W^2 q_{23}^3 (q_{13}-q_{12}) 
\Big[
(d-2) (q_{12}^2
+ q_{13}^2)
+2 q_{12} q_{13} (3 d-8)
\Big]
\\
&&\hspace{-6.5cm}
-M_W^2 q_{12} q_{23} (q_{12}+q_{13})^2 
\Big[(d-2) q_{12}^2+(5 d-18) q_{12} q_{13}+2 q_{13}^2 (d-4)\Big]
\n \\
&&\hspace{-6.5cm}
-M_W^2 q_{23}^2 (q_{12}+q_{13}) 
\Big[
(d-2) (2 q_{12}^3 
-4 q_{12} q_{13}^2
- q_{13}^3)
+(11 d-34) q_{12}^2 q_{13}
\Big]
\n \\
&&\hspace{-6.5cm}
+
q_{13} q_{23}^3 (q_{12}+q_{13})
\Big[(3 d-10) q_{12}^2-(d-2) q_{13} (2 q_{12}- q_{13})\Big]
\n \\
&&\hspace{-6.5cm}
+2 q_{12} q_{13} q_{23}^2 (q_{12}+q_{13})^2 
\Big[(3 d-10) q_{12}-(d-2) q_{13}\Big]
+(d-2) M_W^2 q_{12}^2 q_{13} (q_{12}+q_{13})^3
\n \\
&&\hspace{-6.5cm}
+q_{12}^2 q_{13} q_{23} (q_{12}+q_{13})^3 (3 d-10) 
\Bigg\}C_0(q_{23},0,M_H^2,M_W^2,0,M_W^2)
\n \\
&&\hspace{-6.5cm} +
\dfrac{1}
{4 (d-3) q_{12}^2 q_{13}^2 q_{23}^2}
\Big\{
M_W^4 q_{12}^2 
\Big[(d-2) (q_{13}^2+q_{23}^2)-2 q_{13} q_{23} (d-4)\Big]
\n \\
&&\hspace{-6.5cm}
+2 M_W^4 q_{12} q_{23} (q_{13}+q_{23}) 
\Big[(d-2) q_{23}-(d-4) q_{13}\Big]
\n \\
&&\hspace{-6.5cm}
+2 M_W^2 q_{13} q_{23} (q_{12}+q_{23}) 
\Big[
(d-4) q_{12} q_{13} -
M_H^2 q_{23} (d-2)\Big]
\n \\
&&\hspace{-6.5cm}
+(d-2) q_{23}^2 \Big[ M_W^4 (q_{13}+q_{23})^2
+ q_{13}^2 (q_{12}+q_{23})^2
\Big]
\Big\}D_0(0,0,0,M_H^2;q_{23},q_{13}; M_W^2,0,0,M_W^2), 
\n \\
\n \\
D_{23}(0,0,0,M_H^2;q_{12},q_{13};M_W^2,0,M_W^2,M_W^2)
&=& \n\\
&& \hspace{-6.5cm}
= -\dfrac{1}{q_{13} (q_{13}+q_{23})}
B_0(0,M_W^2,M_W^2)
+ \dfrac{1}{q_{13} q_{23}}
B_0(q_{13},0,M_W^2)
\n \\
&&\hspace{-6.5cm} 
+\dfrac{(q_{23}-q_{13})M_H^2}
{q_{13} q_{23} (q_{13}+q_{23})^2}
B_0(M_H^2,M_W^2,M_W^2)
+
\dfrac{q_{12} (q_{13}-q_{23})-q_{23} (q_{13}+q_{23})}
{q_{13} q_{23} (q_{13}+q_{23})^2}
B_0(q_{12},M_W^2,M_W^2)
\n \\
&&\hspace{-6.5cm} +
\dfrac{1}{4 (d-3) q_{13}^2 q_{23}^2}
\Big\{
 M_W^2 q_{23} q_{13} (d-4)
\n \\
&&\hspace{-6.5cm}
 +(d-2) 
\Big[  M_W^2 q_{23}^2
- M_W^2 q_{12} (q_{13}-q_{23})
+ q_{12} q_{13} (q_{12}+q_{23})
\Big]  
\Big\} C_0(0,0,q_{12},M_W^2,0,M_W^2)
\n \\
&&\hspace{-6.5cm} +
\dfrac{1}{4 (d-3) q_{12} q_{13} q_{23}^2}
\Big\{
M_W^2 q_{12} \Big[(2-d) q_{13}+(3 d - 10) q_{23}\Big]
 \n \\
&&\hspace{-6.5cm}
+M_W^2 q_{23} \Big[(d-4) q_{13}-(d-2) q_{23}\Big]
+ q_{12} q_{13} (q_{12}+q_{23}) (d-2)
\Big\}C_0(0,0,q_{13},0,M_W^2,M_W^2)
\n \\
&&\hspace{-6.5cm} 
+
\dfrac{(q_{12}+q_{23}) 
}{4 (3-d) q_{12} q_{13}^2 q_{23}^2}
\Big\{
M_W^2 q_{23} \Big[(d-2) q_{23}-(d-4) q_{13}\Big]
 \\
&&\hspace{-6.5cm} 
+(2-d) \Big[
M_W^2 q_{12} (q_{13}-q_{23})
- q_{12} q_{13} (q_{12}+q_{23})
\Big]
\Big\}
C_0(0,q_{13},M_H^2,M_W^2,0,M_W^2)
\n \\
&&\hspace{-6.5cm} +
\dfrac{1}{4 (d-3) q_{12} 
q_{13}^2 q_{23}^2 (q_{13}+q_{23})}
\Big\{
M_W^2 (q_{13}+q_{23}) 
\Big[
(d-2) [q_{12} (q_{13}^2
+ q_{23}^2 )
+ q_{23}^3 ]
 \n \\
&&\hspace{-6.5cm}
-(d-4) q_{13} q_{23} (2 q_{12}+q_{13})
+2 q_{13} q_{23}^2 \Big]
-q_{12} q_{13} 
\Big[q_{12} [(d-2) q_{13} (q_{13}+2 q_{23})
+(10-3 d) q_{23}^2]
\n \\
&&\hspace{-6.5cm}
+q_{23} (q_{13}+q_{23}) 
[(d-2) q_{13}+(10-3 d) q_{23}]\Big]
\Big\}C_0(q_{12},0,M_H^2,M_W^2,M_W^2,M_W^2)
\n \\
&&\hspace{-6.5cm}
+
\dfrac{1}
{4 (3-d) q_{12} q_{13}^2 q_{23}^2}
\Big\{
M_W^4 
\Big[
(d-2) q_{12} 
q_{13}^2+q_{23}^2 [(d-2) q_{12}+2 q_{13}]
\n \\
&&\hspace{-6.5cm}
-(d-4) q_{13} q_{23} 
(2 q_{12}+q_{13})+(d-2) q_{23}^3
\Big]
\n \\
&&\hspace{-6.5cm}
-2 M_W^2 q_{12} q_{13} 
\Big[
(d-2) q_{12} (q_{13}-q_{23})
+q_{23} [q_{13}+(2-d) q_{23}]
\Big]
\n \\
&&\hspace{-6.5cm}
+(d-2) q_{12}^2 q_{13}^2 (q_{12}+q_{23})
\Big\}
D_0(0,0,0,M_H^2; q_{12},q_{13}; M_W^2,0,M_W^2,M_W^2), 
\n \\
\n\\
D_{23}(0,0,0,M_H^2;q_{23},q_{13};M_W^2,0,0,M_W^2)
&=& \n \\
&& \hspace{-6.5cm}
-\dfrac{1}{q_{13} (q_{12}+q_{13})}
B_0(0,0,M_W^2)
+
\dfrac{1}{q_{12} q_{13}}
B_0(q_{13},0,M_W^2)
\n \\
&&\hspace{-6.5cm} +
\dfrac{M_H^2 (q_{12}-q_{13})}
{q_{12} q_{13} (q_{12}+q_{13})^2}
B_0(M_H^2,M_W^2,M_W^2)
+
\dfrac{(q_{13} q_{23} - M_H^2 q_{12})}
{q_{12} q_{13} (q_{12}+q_{13})^2}
B_0(q_{23},M_W^2,0)
\n \\
&&\hspace{-6.5cm} +
\dfrac{1}{4 (d-3) q_{12}^2 q_{13} q_{23}}
\Big\{
M_W^2 q_{12} \Big[(d-4) q_{13}-(d-2) q_{23}\Big]
\n \\
&&\hspace{-6.5cm}
-(d-2) q_{23} \Big[
M_W^2 (q_{13}+q_{23})
- q_{13} (q_{12}+q_{23})\Big]
\Big\}C_0(0,0,q_{13},0,0,M_W^2)
\n \\
&&\hspace{-6.5cm} +
\dfrac{1}{4 (d-3) q_{12}^2 q_{13}^2}
\Big\{
M_W^2 q_{12} 
\Big[(d-4) q_{13}-(d-2) q_{23}\Big]
\n \\
&&\hspace{-6.5cm}
-(d-2) q_{23} \Big[
M_W^2 (q_{13}+q_{23})
- q_{13} (q_{12}+q_{23})\Big]
\Big\}C_0(0,0,q_{23},M_W^2,0,0)
\n \\
&&\hspace{-6.5cm} +
\dfrac{(q_{12}+q_{23})}
{4 (d-3) q_{12}^2 q_{13}^2 q_{23}} 
\Big\{
M_W^2 q_{12} \Big[(d-4) q_{13}+(d-2) q_{23}\Big]
\n \\
&&\hspace{-6.5cm}
+(d-2) q_{23} \Big[
M_W^2 (q_{13}+q_{23})
- q_{13} (q_{12}+q_{23})\Big]
\Big\}C_0(0,q_{13},M_H^2,M_W^2,0,M_W^2)
\\
&&\hspace{-6.5cm} 
+
\dfrac{1}
{4 (3-d) q_{12}^2 q_{13}^2 q_{23} (q_{12}+q_{13})^2}
\Big\{
(d-4) M_W^2 q_{12} q_{13} (q_{12}+q_{13})^3
\n \\
&&\hspace{-6.5cm} 
+
M_W^2 q_{23}^2 (q_{12}-q_{13}) 
\Big[
(d-2) (q_{12}^2 + q_{13}^2)
+2 q_{12} q_{13} (3 d-8)
\Big]
\n \\
&&\hspace{-6.5cm} 
+M_W^2 q_{23} (q_{12}+q_{13}) \Big[
(d-2) (q_{12}^3
-3 q_{12} q_{13}^2
- q_{13}^3)
+(7 d-22) q_{12}^2 q_{13}
\Big]
\n \\
&&\hspace{-6.5cm} 
+
q_{13} q_{23}^2 (q_{12}+q_{13})
\Big[(10-3 d) q_{12}^2
+(d-2) q_{13} (2 q_{12}
+ q_{13})
\Big]
\n \\
&&\hspace{-6.5cm} 
+ q_{12} q_{13} q_{23} (q_{12}+q_{13})^2 \Big[
(10-3 d) q_{12}+(d-2) q_{13}\Big]
\Big\} C_0(q_{23},0,M_H^2,M_W^2,0,M_W^2)
\n \\
&&\hspace{-6.5cm} 
+  
\dfrac{q_{13} q_{23} - M_W^2 (q_{13}+q_{23})}
{4 (d-3) q_{12}^2 q_{13}^2 q_{23}}
\Big\{M_W^2 q_{12} \Big[(d-2) q_{23}-(d-4) q_{13}\Big]
\n \\
&&\hspace{-6.5cm} 
+ (d-2) q_{23} \Big[ M_W^2 (q_{13}+q_{23})
- q_{13} (q_{12}+q_{23}) \Big]\Big\}
D_0(0,0,0,M_H^2;q_{23},q_{13}; M_W^2,0,0,M_W^2), 
\n \\
\n \\
D_{23}(0,0,0,M_H^2;q_{23},q_{12};M_W^2,M_W^2,0,M_W^2) 
&=&
\n \\
&&\hspace{-6.5cm} 
-\dfrac{1}{q_{12} (q_{12}+q_{13})}
 B_0(0,0,M_W^2)
 +
\dfrac{1}{q_{12} q_{13}}
B_0(q_{12},M_W^2,M_W^2)
\n \\
&&\hspace{-6.5cm} 
+
\dfrac{M_H^2 (q_{13}-q_{12}) 
}{q_{12} q_{13} (q_{12}+q_{13})^2}
B_0(M_H^2,M_W^2,M_W^2)
-\dfrac{q_{12} (q_{13} 
- q_{23})+q_{13} (q_{13}+q_{23})}
{q_{12} q_{13} (q_{12}+q_{13})^2}
B_0(q_{23},M_W^2,0)
\n \\
&&\hspace{-6.5cm} +
\dfrac{1}{4 (d-3) q_{12} q_{13}^2 q_{23}}
\Big\{
M_W^2 
\Big[
(d-2) (q_{13}^2 - q_{23}^2)
-4 q_{13} q_{23} (d-3)
\Big]
\n \\
&&\hspace{-6.5cm}
+
(d-2) q_{12} q_{23} (q_{13}+q_{23})\Big\}
C_0(0,0,q_{12},M_W^2,0,M_W^2)
\n \\
&&\hspace{-6.5cm} +
\dfrac{(d-2) (q_{13}+q_{23})}
{4 (3-d) q_{12}^2 q_{13}^2} 
\Big[M_W^2 (q_{13}+q_{23})-q_{12} q_{23}\Big] 
C_0(0,0,q_{23},M_W^2,M_W^2,0)
\n \\
&&\hspace{-6.5cm} 
+
\dfrac{(d-2) (q_{13}+q_{23})^2 }
{4 (d-3) q_{12}^2 q_{13}^2 q_{23}}
\Big[M_W^2 (q_{13}+q_{23})-q_{12} q_{23}\Big]
C_0(0,q_{12},M_H^2,M_W^2,M_W^2,M_W^2)
\n \\
&&\hspace{-6.5cm} +
\dfrac{1}
{4 (3-d) q_{12}^2 q_{13}^2 q_{23} (q_{12}+q_{13})^2}
\Bigg\{
M_W^2 q_{12}^3 
\Big[
(d-2) (q_{13}^2
- q_{23}^2)
-4 q_{13} q_{23} (d-3)
\Big]
\\
&&\hspace{-6.5cm} 
+(d-2) M_W^2 q_{13}^3 (q_{13}+q_{23})^2
+M_W^2 q_{12}^2 q_{13} 
\Big[
 (d-2) q_{13} (3 q_{13}
+2 q_{23})
+(14-5 d) q_{23}^2
\Big]
\n \\
&&\hspace{-6.5cm} 
+M_W^2 q_{12} q_{13}^2 (q_{13}+q_{23}) 
\Big[3 q_{13} (d-2)+(5 d-14) q_{23}\Big]
\n \\
&&\hspace{-6.5cm} 
+q_{12} q_{23} (q_{12}+q_{13}) 
\Big[
(d-2) q_{12}^2 (q_{13}+q_{23})
+2 q_{12} q_{13} 
\Big[(d-2) q_{23}-(d-4) q_{13}\Big]
\n \\
&&\hspace{-2.0cm} 
-(3 d-10) q_{13}^2 (q_{13}+q_{23})
\Big]
\Bigg\} C_0(q_{23},0,M_H^2,M_W^2,0,M_W^2)
\n \\
&&\hspace{-6.5cm} +
\dfrac{(q_{13}+q_{23})}
{4 (3-d) q_{12}^2 q_{13}^2 q_{23}} 
\Big\{
(d-2) M_W^4 (q_{13}+q_{23})^2
-2 M_W^2 q_{12} q_{23} \Big[(d-4) q_{13}+(d-2) q_{23}\Big]
\n \\
&&\hspace{-6.5cm}
+(d-2) q_{12}^2 q_{23}^2
\Big\} D_0(0,0,0,M_H^2;q_{23},q_{12}; M_W^2,M_W^2,0,M_W^2), 
\n \\
\n \\
D_{33}(0,0,0,M_H^2;q_{23},q_{12};M_W^2,M_W^2,0,M_W^2)
&=&\\
&&
\hspace{-6.5cm}
=
\dfrac{1}{q_{12} (q_{12}+q_{13})}
B_0(0,0,M_W^2)
-\dfrac{1}{q_{12} q_{13}}
B_0(q_{12},M_W^2,M_W^2) 
\n \\
&&\hspace{-6.5cm} +
\dfrac{(M_H^2 q_{12}-q_{13} q_{23}) 
}{q_{12} q_{13} (q_{12}+q_{13})^2}
B_0(M_H^2,M_W^2,M_W^2)
-\dfrac{q_{23} (q_{12}-q_{13}) 
}{q_{12} q_{13} (q_{12}+q_{13})^2}
B_0(q_{23},M_W^2,0)
\n \\
&&\hspace{-6.5cm} +
\dfrac{1}{4 (d-3) q_{12} q_{13}^2}
\Big\{
M_W^2 \Big[(3 d-10) q_{13}+(d-2) q_{23}\Big]
\n \\
&&\hspace{-3cm} 
-(d-2) q_{12} q_{23}
\Big\}C_0(0,0,q_{12},M_W^2,0,M_W^2)
\n \\
&&\hspace{-6.5cm} +
\dfrac{(d-2) q_{23}}{4 (d-3) q_{12}^2 q_{13}^2}
\Big[M_W^2 (q_{13}+q_{23})-q_{12} q_{23}\Big]
C_0(0,0,q_{23},M_W^2,M_W^2,0)
\n \\
&&\hspace{-6.5cm} +
\dfrac{(d-2) (q_{13}+q_{23})}{4 (3-d) q_{12}^2 q_{13}^2} 
\Big[M_W^2 (q_{13}+q_{23})-q_{12} q_{23}\Big]
C_0(0,q_{12},M_H^2,M_W^2,M_W^2,M_W^2)
\n \\
&&\hspace{-6.5cm} +
\dfrac{1}
{4 (d-3) q_{12}^2 q_{13}^2 (q_{12}+q_{13})^2}
\Bigg\{
M_W^2 q_{12}^3 \Big[(10-3 d) q_{13}-(d-2) q_{23}\Big]
\n \\
&&\hspace{-6.5cm}
+M_W^2 q_{12}^2 q_{13} \Big[(14-5 d) q_{23}-(d-6) q_{13}\Big]
\\
&&\hspace{-6.5cm}
+M_W^2 q_{12} q_{13}^2 \Big[3 q_{13} (d-2)+(5 d-14) q_{23}\Big]
+(d-2) M_W^2 q_{13}^3 (q_{13}+q_{23})
\n \\
&&\hspace{-6.5cm}
+q_{12} q_{23} (q_{12}+q_{13}) 
\Big[
(d-2) q_{12} (q_{12}
+2 q_{13})
+(10-3 d) q_{13}^2
\Big]
\Bigg\} C_0(q_{23},0,M_H^2,M_W^2,0,M_W^2)
\n \\
&&\hspace{-6.5cm} +
\dfrac{1}
{4 (d-3) q_{12}^2 q_{13}^2}
\Big\{
(d-2) \Big[ M_W^4 (q_{13}+q_{23})^2
+ q_{12}^2 q_{23}^2 \Big]
\n \\
&&\hspace{-6.5cm} 
-2 M_W^2 q_{12} q_{23} \Big[(d-4) q_{13}+(d-2) q_{23}\Big]
\Big\}D_0(0,0,0,M_H^2;q_{23},q_{12}; M_W^2,M_W^2,0,M_W^2),
\n \\
\n \\
D_{33}(0,0,0,M_H^2;q_{12},q_{13};M_W^2,0,M_W^2,M_W^2)
&=&\n \\
&& \hspace{-6.5cm} =
\dfrac{1}{q_{13}(q_{13} + q_{23})} 
B_0(0,M_W^2,M_W^2)
-\dfrac{1}{q_{13} q_{23}}
B_0(q_{13},0,M_W^2)
\n \\
&&
\hspace{-6.5cm} +
\dfrac{q_{12} (q_{13}-q_{23})+q_{13} (q_{13}+q_{23})}
{q_{13} q_{23} (q_{13}+q_{23})^2}
B_0(M_H^2,M_W^2,M_W^2)
+
\dfrac{q_{12} (q_{23}-q_{13})}{q_{13} q_{23} (q_{13}+q_{23})^2}
B_0(q_{12},M_W^2,M_W^2)
\n \\
&&\hspace{-6.5cm}  
+\dfrac{q_{12} (2-d) [M_W^2 (q_{23}
- q_{13})+q_{12} q_{13}]}{4 (d-3) q_{13}^2 q_{23}^2}
C_0(0,0,q_{12},M_W^2,0,M_W^2)
\n \\
&&\hspace{-6.5cm} +
\dfrac{q_{13} (M_W^2-q_{12}) (d-2) 
+ M_W^2 q_{23}(10-3 d)}{4 (d-3) q_{13} q_{23}^2}
C_0(0,0,q_{13},0,M_W^2,M_W^2)
 \\
&&\hspace{-6.5cm} +
\dfrac{1}{4 (d-3) q_{13}^2 q_{23}^2}
\Big\{
M_W^2 q_{13} q_{23} (10-3 d)
 \\
&&\hspace{-6.5cm}
+(d-2) \Big[
M_W^2 q_{23}^2
- M_W^2 q_{12} (q_{13}-q_{23})
+ q_{12} q_{13} (q_{12}+q_{23})
\Big]
\Big\} C_0(0,q_{13},M_H^2,M_W^2,0,M_W^2)
\n \\
&&\hspace{-6.5cm} +
\dfrac{1}
{4 (d-3) q_{13}^2 q_{23}^2 (q_{13}+q_{23})}
\Big\{
q_{12} q_{13} \Big[
q_{13} ( q_{13}+2 q_{23})(d-2) +(10-3 d) q_{23}^2
\Big]
\n \\
&&\hspace{-6.cm}
-M_W^2 (q_{13}+q_{23}) \Big[
(d-2) (q_{13}^2
+ q_{23}^2)
-2 (d-4) q_{13} q_{23}
\Big]
\Big\}C_0(q_{12},0,M_H^2,M_W^2,M_W^2,M_W^2)
\n \\
&&\hspace{-6.5cm} +
\dfrac{1}
{4 (d-3) q_{13}^2 q_{23}^2}
\Big\{
(d-2) \Big[
M_W^4 q_{23}^2+ q_{13}^2 (M_W^2-q_{12})^2
\Big]
\n \\
&&\hspace{-6.5cm}
+2 M_W^2 q_{13} q_{23} \Big[
(d-2) q_{12}-(d-4) M_W^2
\Big]
\Big\}
 D_0(0,0,0,M_H^2; q_{12},q_{13}; M_W^2,0,M_W^2,M_W^2),
\n \\
\n \\
D_{33}(0,0,0,M_H^2;q_{23},q_{13};M_W^2,0,0,M_W^2) 
&=&
\n \\
&& \hspace{-6.5cm} =
\dfrac{1}{q_{13} (q_{12}+q_{13})}
B_0(0,0,M_W^2)
-\dfrac{1}{q_{12} q_{13}}
B_0(q_{13},0,M_W^2)
\n \\
&&\hspace{-6.5cm} +
\dfrac{1}{q_{12} q_{13} (q_{12}+q_{13})^2}
\Big[q_{12} (q_{13}-q_{23})+q_{13} (q_{13}+q_{23})\Big]
B_0(M_H^2,M_W^2,M_W^2)
\n \\
&&\hspace{-6.5cm}
+
\dfrac{q_{23} (q_{12}-q_{13})}{q_{12} q_{13} (q_{12}+q_{13})^2}
B_0(q_{23},M_W^2,0)
\n \\
&&\hspace{-6.5cm} +
\dfrac{(d-2)}{4 (d-3) q_{12}^2 q_{13}} 
\Big[M_W^2 (q_{13}+q_{23})-q_{13} q_{23}\Big]
C_0(0,0,q_{13},0,0,M_W^2)\n \\
&&\hspace{-6.5cm} +
\dfrac{(d-2) q_{23} }{4 (d-3) q_{12}^2 q_{13}^2}
\Big[M_W^2 (q_{13}+q_{23})-q_{13} q_{23}\Big]
C_0(0,0,q_{23},M_W^2,0,0)
\n \\
&&\hspace{-6.5cm} +
\dfrac{1}{4 (d-3) q_{12}^2 q_{13}^2}
\Big\{
(d-2) q_{13} q_{23} (q_{12}+q_{23})
-M_W^2 q_{12} \Big[(3 d-10) q_{13}+(d-2) q_{23}\Big]
\\
&&\hspace{-6.5cm}
-M_W^2 q_{23} (q_{13}+q_{23}) (d-2)
\Big\}C_0(0,q_{13},M_H^2,M_W^2,0,M_W^2)
\n \\
&&\hspace{-6.5cm} +
\dfrac{1}{4 (d-3) q_{12}^2 q_{13}^2 (q_{12}+q_{13})^2}
\Big\{
M_W^2 q_{12}^3 \Big[(3 d-10) q_{13}+(d-2) q_{23}\Big]
\n \\
&&\hspace{-6.5cm}
+M_W^2 q_{12}^2 q_{13} \Big[(d-6) q_{13}+(5 d-14) q_{23}\Big]
\n \\
&&\hspace{-6.5cm}
+M_W^2 q_{12} q_{13}^2 \Big[(14-5 d) q_{23}-3 (d-2) q_{13}\Big]
-(d-2) M_W^2 q_{13}^3 (q_{13}+q_{23})
\n \\
&&\hspace{-6.5cm}
-q_{13} q_{23} (q_{12}+q_{13}) 
\Big[(3 d-10) q_{12}^2-2 (d-2) q_{12} q_{13}-(d-2) q_{13}^2\Big]
\Big\}\times 
\n\\
&& \hspace{-6.5cm}
\times C_0(q_{23},0,M_H^2,M_W^2,0,M_W^2)
\n \\
&&\hspace{-6.5cm} +
\dfrac{(d-2)}{4 (d-3) q_{12}^2 q_{13}^2}
\Big[q_{13} q_{23}-M_W^2 (q_{13}+q_{23})\Big]^2
D_0(0,0,0,M_H^2;q_{23},q_{13}; M_W^2,0,0,M_W^2). \n
\end{eqnarray}
\section{Expansion for form factors of 
    $\mathcal{A}^{(W)}_{\text{tri}}$ 
and $\mathcal{A}^{(f)}_{\text{tri}}$ }         
Since the dominant contributions of triangle diagrams
with $W$ boson and fermion internal lines to the decay 
channels. It is worth to perform the $\varepsilon$-expansion 
for the form factors of $\mathcal{A}^{(W)}_{\text{tri}}$ 
and $\mathcal{A}^{(f)}_{\text{tri}}$. Both 
representations of form factors give same results
\begin{eqnarray}
\mathcal{A}^{(W)}_{\text{tri}} &=& 
\dfrac{\alpha ^2
\Big[ (M_H^2-q_{12}) g^{\mu \nu} -2 q^{\nu } q_3^\mu \Big]
}{4 M_H^2 M_W^3 s_W^3 (M_H^2 - q_{12})^2 (q_{12} -M_Z^2 + i \Gamma_Z M_Z)}
\times
\n \\
&&\times
\Bigg\{
M_H^2 M_W^2 
\Big[M_H^2 (q_{12}-6 M_W^2)+12 M_W^4+6 M_W^2 q_{12}-2 q_{12}^2\Big]
\times
\n \\
&&\hspace{3cm} \times
 \ln ^2 \Bigg[
 \dfrac{-M_H^2+\sqrt{M_H^4-4 M_H^2 M_W^2}+2 M_W^2}{2 M_W^2}\Bigg]
 \n \\
&&
 +M_H^2 
 \Big[2 M_H^4 M_W^2-M_H^4 q_{12}+12 M_H^2 M_W^4-4 M_H^2 M_W^2 q_{12}
 \n \\
&&\hspace{0cm}
 +M_H^2 q_{12}^2-12 M_W^4 q_{12}+2 M_W^2 q_{12}^2
\Big] \n \\
&&\hspace{0cm}
+M_H^2\sqrt{q_{12}^2-4 M_W^2 q_{12}} 
\Big[M_H^2 (q_{12}-2 M_W^2)+2 M_W^2 (q_{12}-6 M_W^2)\Big] 
\times
\n \\
&&\hspace{2.75cm} \times
\ln \Bigg[\dfrac{-q_{12} + \sqrt{q_{12}^2-4 M_W^2 q_{12}}+2 M_W^2}{2 M_W^2}\Bigg]
\\
&&
-M_W^2M_H^2 \Big[M_H^2 (q_{12}-6 M_W^2)+12 M_W^4+6 M_W^2 q_{12}-2 q_{12}^2\Big]
\times
\n \\
&&\hspace{3cm} \times
 \ln ^2 \Bigg[\dfrac{-q_{12} + \sqrt{q_{12}^2-4 M_W^2 q_{12}}+2 M_W^2}{2 M_W^2}\Bigg]
 \n \\
&&
+q_{12} \sqrt{M_H^4-4 M_H^2 M_W^2} 
\Big[M_H^2 (2 M_W^2-q_{12})+12 M_W^4-2 M_W^2 q_{12}\Big]
\times
\n \\
&&\hspace{3cm} \times
 \ln \Bigg[\dfrac{-M_H^2+\sqrt{M_H^4-4 M_H^2 M_W^2}+2 M_W^2}{2 M_W^2}\Bigg]
\Bigg\} \times
\n\\
&& \times 
\bar{u}(q_1) \gamma ^\mu P_L v(q_2) \epsilon^{*}_{\nu}(q_3), 
\n\\
&& \n\\
 \mathcal{A}^{(f)}_{\text{tri}} &=& 
\dfrac{\alpha ^2 m_f^2 N_C^f Q_f
\Big[2 q^\nu q_3^\mu +(q_{12}-M_H^2) g^{\mu \nu} \Big]
(T^3_f - 2 Q_f s_W^2)
 }{4 M_H^2 M_W c_W^2 s_W^3 (M_H^2-q_{12})^2
(q_{12} -M_Z^2 + i \Gamma_Z M_Z)}
 \times
 \n \\
 &&\hspace{-0.5cm} \times
 \Bigg\{
M_H^2 \Big[
(4 m_f^2-M_H^2+q_{12}) 
\ln ^2 \Bigg[\dfrac{-q_{12} + \sqrt{q_{12}^2 - 4 q_{12} m_f^2}+2 m_f^2}{2 m_f^2}\Bigg]
 \n \\
 &&\hspace{0.05cm} 
 +4 \sqrt{q_{12}^2 - 4 q_{12} m_f^2} 
 \ln \Bigg[\dfrac{-q_{12} + \sqrt{q_{12}^2 - 4 q_{12} m_f^2}+2 m_f^2}{2 m_f^2}\Bigg]
 -4 M_H^2 +4 q_{12}
 \Big]
\\
 &&\hspace{0.05cm}  
 +M_H^2 (M_H^2 -4 m_f^2-q_{12}) 
 \ln ^2 \Bigg[\dfrac{-M_H^2 + \sqrt{M_H^4-4 m_f^2 M_H^2}+2 m_f^2}{2 m_f^2}\Bigg]
\n \\
 &&\hspace{0.05cm}  
 -4 q_{12} \sqrt{M_H^4-4 M_H^2 m_f^2} 
 \ln \Bigg[\dfrac{-M_H^2 + \sqrt{M_H^4-4 m_f^2 M_H^2}+2 m_f^2}{2 m_f^2}\Bigg]
 \Bigg\}\times 
 \n\\
 && \times
 \bar{u}(q_1) \gamma ^\mu P_L v(q_2)\epsilon^{*}_{\nu}(q_3). 
 \n
\end{eqnarray}
For $W$ bosons and fermions exchanging in the loop, their masses are 
included the Feynman's prescription as $M_W^2 -i\rho$ 
($m_f^2 -i\rho$ respectively). 
As a result, all logarithmic functions in the above equations 
are well-defined in complex plane.
\section{Decay rate} 
Differential decay width is given by 
\begin{eqnarray}
 \dfrac{d^2\Gamma(H\rightarrow 
\nu_l\bar{\nu}_l  \gamma)}{dq_{12}dq_{13}}
 = \dfrac{1}{256\pi^3 \; M_H^3}
 \sum\limits_{\text{spin}}|\mathcal{A}_{\text{tri}}
+\mathcal{A}_{\text{box}} |^2, 
\end{eqnarray}
We have used following notations:
\begin{eqnarray}
\mathcal{A}_{\text{tri}} &=& 
\Big[F_{1} q_3^{\mu}q^{\nu} +  
F_{2} g^{\mu\nu} \Big] 
\bar{u}(q_1)\gamma^{\mu}\; 
P_L v(q_2) \epsilon^{*}_{\nu}(q_3),\\
\mathcal{A}_{\text{box}} 
&=&  
\bar{u}(q_1)
\Big\{ 
F_{3}q_1^{\nu} \slashed{q_3}
+  F_{4}q_2^{\nu} \slashed{q_3} 
+ F_{5}
\gamma^{\nu}\Big\}
P_L v(q_2) \epsilon^{*}_{\nu}(q_3).
\end{eqnarray}
After performing 
spin-sum for the above amplitude-squared, one gets 
several representations (together with Eq.~\ref{decay1}) 
for the differential decay width as follows: 
\begin{eqnarray}
\dfrac{d^2\Gamma(H\rightarrow 
\nu_l\bar{\nu}_l  \gamma)}{dq_{12}dq_{13}}
 &=& \dfrac{q_{12}}{256\pi^3 M_H^3 (q_{13}+q_{23})^2}
\times
\n \\
&&\hspace{-3.5cm} \times
\Big\{
2 (q_{13}^2 + q_{23}^2) |F_2|^2 
+ q_{13}^2 (q_{13}+q_{23})^2 |F_3|^2
+ 2 (q_{13}+q_{23})^2 |F_5|^2
\\
&&\hspace{-2.5cm} 
+ 4 q_{23} (q_{13} + q_{23}) \, \text{Re} [F_2 F_5^*]
+ 2 q_{13} (q_{23}^2 - q_{13}^2) \, \text{Re} [F_2 F_3^*]
+ 2 q_{13} (q_{13}+q_{23})^2 \, \text{Re} [F_3 F_5^*]
\Big\}
,\n \\
\dfrac{d^2\Gamma(H\rightarrow 
\nu_l\bar{\nu}_l  \gamma)}{dq_{12}dq_{13}}
 &=& \dfrac{q_{12}}{256\pi^3 M_H^3 (q_{13}+q_{23})^2}
\times
\n \\
&&\hspace{-3.5cm} \times
\Big\{
2 (q_{13}^2 + q_{23}^2) |F_2|^2 
+ q_{23}^2 (q_{13}+q_{23})^2 |F_4|^2
+ 2 (q_{13}+q_{23})^2 |F_5|^2
\\
&&\hspace{-2.5cm} 
+ 4 q_{13} (q_{13} + q_{23}) \, \text{Re} [F_2 F_5^*]
+ 2 q_{23} (q_{13}^2 - q_{23}^2) \, \text{Re} [F_2 F_4^*]
+ 2 q_{23} (q_{13}+q_{23})^2 \, \text{Re} [F_4 F_5^*]
\Big\}.
\n
 \end{eqnarray}

\end{document}